%% file: SynAct.tex
\documentclass[10pt,journal]{IEEEtran}
\IEEEoverridecommandlockouts 
\input{setting-ieee}
\graphicspath{{./figs/}{../}}

\newif\ifarxivversion
\arxivversiontrue
\begin{document}

\twocolumn

\title{
    SynAct: A Reasoning-Acting Large Language Model Agent for Adaptive Synthesis Optimization
}

\author{
    Fangzhou Liu, \quad
    Peiyi Han,    \quad
    Jiawei Liu,   \quad
    Yuan Pu,      \quad
    Zhuolun He,   \quad
    Rongliang Fu, \quad
    Tsung-Yi Ho,  \quad
    Bei Yu \\
    \thanks{
        Fangzhou Liu, Peiyi Han, Jiawei Liu, Yuan Pu, Zhuolun He, Rongliang Fu, Tsung-Yi Ho, and Bei Yu are with the Department of Computer Science and Engineering, The Chinese University of Hong Kong, Hong Kong SAR.
    }
    \thanks{
        Bei Yu and Rongliang Fu are the corresponding authors (e-mail: byu@cse.cuhk.edu.hk; rlfu@cse.cuhk.edu.hk).
    }
    \ifarxivversion
    \thanks{
        This work has been submitted to the IEEE for possible publication. Copyright may be transferred without notice, after which this version may no longer be accessible.
    }
    \fi
}

\maketitle
\thispagestyle{plain}
\pagestyle{plain}

\input{doc/abstract}
\input{doc/intro}
\input{doc/prelim}
\input{doc/algo}
\input{doc/exp}
\input{doc/conclu}
\input{doc/acknowledge}

{
\bibliographystyle{IEEEtran}
\bibliography{ref/Top,ref/reference}
}

\input{doc/bio}

\end{document}

%% file: setting-ieee.tex
\usepackage{blkarray}                                      
\usepackage{graphicx}                                      
\usepackage{amsmath}
\usepackage{amssymb}
\usepackage{amsfonts}
\usepackage{amsthm}
\usepackage[mathcal]{eucal}
\usepackage{mathrsfs}
\usepackage{booktabs}
\usepackage{enumerate}
\usepackage{multirow}
\usepackage[subrefformat=parens,farskip=0pt,justification=centering]{subfig}
\usepackage{color}
\usepackage{cite}                                          
\usepackage{comment}                                       
\usepackage{soul}                                          
\soulregister\cite7
\soulregister\ref7
\soulregister\pageref7
\usepackage{etoolbox}                                      
\usepackage{url}
\usepackage{nth}                                           
\usepackage{bm}                                            
\usepackage{courier}
\usepackage{balance}
\usepackage{threeparttable}
\usepackage{xcolor,colortbl}
\usepackage{footnote}
\usepackage{listings}
\usepackage{setspace}                                      
\usepackage[inline]{enumitem}
\usepackage{verbatim}
\usepackage[bookmarks=false]{hyperref}
\hypersetup{
    colorlinks = true,
    citecolor  = blue,
    linkcolor  = blue,
    urlcolor   = blue,
}
\usepackage{tikz}
\usetikzlibrary{patterns,snakes}
\usetikzlibrary{positioning,calc,fit,decorations.pathmorphing,shapes.geometric, shapes.gates.logic.US, calc}
\usetikzlibrary{arrows,arrows.meta,decorations.markings,shapes,shapes.arrows}
\usetikzlibrary{decorations,decorations.pathreplacing}
\usetikzlibrary{backgrounds}
\usepackage{filecontents}                                  
\usepackage{pgfplots}
\usepackage{pgfplotstable}
\usepgfplotslibrary{groupplots}
\usepackage{scalefnt}
\pgfplotsset{compat=newest}
\usepackage{caption}
\usepackage{pifont}                                        
\usepackage{cleveref}
\Crefformat{figure}{Fig.~#2#1#3}                           
\Crefname{subfigure}{Fig.}{Figs.}
\Crefname{figure}{Fig.}{Figs.}
\Crefformat{table}{TABLE~#2#1#3}                           
\usepackage[figuresright]{rotating}

\usepackage{algorithm}
\iftrue
\usepackage{algpseudocode}                                 
\algrenewcommand\textproc{\texttt}
\makeatletter
\let\OldStatex\Statex
\renewcommand{\Statex}[1][3]{%
  \setlength\@tempdima{\algorithmicindent}%
  \OldStatex\hskip\dimexpr#1\@tempdima\relax
}
\makeatother
\else
\usepackage{algorithmic}
\fi

\definecolor{CUHKorange}{RGB}{244,106,18} 
\definecolor{CUHKblue}{RGB}{0,111,190}    
\definecolor{CUHKgreen}{RGB}{0,127,128}   
\definecolor{CUHKred}{RGB}{228,46,36}     
\definecolor{CUHKyellow}{RGB}{198,148,34} 
\definecolor{CUHKdark}{RGB}{114,44,114}   
\definecolor{CUHKmiddle}{RGB}{144,44,144} 
\definecolor{CUHKlight}{RGB}{167,44,167} 
\definecolor{CUHKpurple}{RGB}{117,15,109}
\definecolor{CUHKgold}{RGB}{221,163,0}
\definecolor{CUHKribbon}{RGB}{244,223,176}
\definecolor{CUHKblack}{RGB}{34,24,21}

\newcommand{\tool}[1]{$\mathsf{#1}$}
\newcommand{\altisyn}{AltiSyn\textsuperscript{\textregistered}}
\renewcommand{\bf}[1]{\textbf{#1}}
\renewcommand{\tt}[1]{\texttt{#1}}

\newcommand{\minisection}[1]{\vspace{.1in}\noindent{\textbf{#1}}.}

\usepackage{tcolorbox}
\tcbuselibrary{skins,breakable}
    {\endtcolorbox}
    {\endtcolorbox}
\newenvironment{promptbox}{%
    \begin{tcolorbox}[breakable,
    boxsep=2pt,left=4pt,right=4pt,top=3pt,bottom=3pt,
    colback=black!3,colframe=black!20]}%
    {\end{tcolorbox}}

\usepackage[top=0.80in,bottom=0.88in,left=0.63in,right=0.63in]{geometry}
\newtheorem{mydefinition}{\textbf{Definition}}

\crefname{mytheorem}{Theorem}{Theorems}
\crefname{mylemma}{Lemma}{Lemmas}
\crefname{myclaim}{Claim}{Claims}
\crefname{myproperty}{Property}{Properties}
\crefname{mycorollary}{Corollary}{Corollaries}

\RequirePackage[normalem]{ulem} 
\RequirePackage{color}\definecolor{RED}{rgb}{1,0,0}\definecolor{BLUE}{rgb}{0,0,1} 

%% file: doc/abstract.tex
\begin{abstract}
    Logic synthesis transforms RTL designs into gate-level netlists, where PPA results are highly sensitive to the choice of optimization commands, making synthesis tuning both high-dimensional and expensive.
    Previous approaches fall into two categories: automated methods, which perform black-box search over fixed action spaces with limited decision-level interpretability, and LLM-based methods, which typically generate static scripts upfront and cannot adapt to evolving circuit states.
    We present SynAct, an adaptive closed-loop LLM reasoning--acting agent that iteratively diagnoses live synthesis reports and reasons over the current circuit state, retrieved tool knowledge, and historical optimization experience to issue targeted commands.
    SynAct focuses on improving timing, particularly worst negative slack (WNS), while maintaining balanced area and power trade-offs.
    Experiments on a commercial synthesis tool across 14 designs show that SynAct reduces average WNS to 27\% of that from bootstrap synthesis.
\end{abstract}


%% file: doc/intro.tex
\section{Introduction}

\begin{figure}[tb!]
    \centering
    \includegraphics[width=0.49\textwidth]{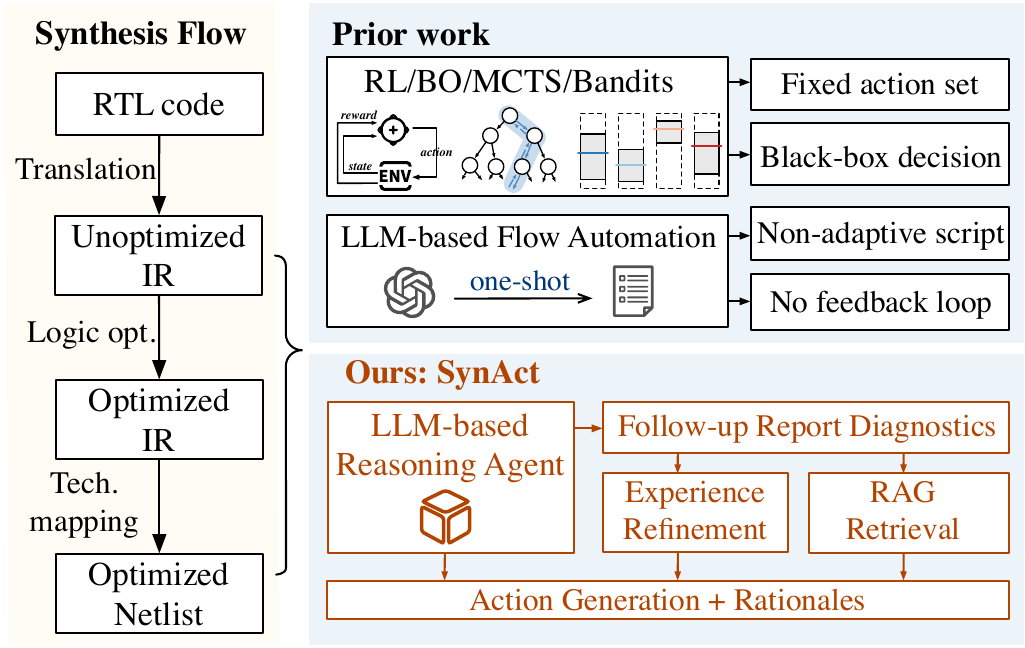}
    \caption{Traditional synthesis flow vs.\ our LLM-based agent SynAct. ``IR'' denotes intermediate representation.}
    \label{fig:comparison}
\end{figure}

\IEEEPARstart{L}{ogic} synthesis transforms RTL designs into gate-level netlists.
As shown in \Cref{fig:comparison}, a typical flow comprises translation, logic optimization, and technology mapping, where the latter two stages involve extensive algorithmic choices that commercial tools encapsulate into large command sets with diverse configurable parameters.
Prior studies show that the choice and ordering of commands significantly affect PPA outcomes~\cite{yu2018developing}, making synthesis tuning a high-dimensional and high-cost design-space exploration challenge.
Among the PPA metrics, timing is particularly critical, as violations can necessitate flow re-tuning and additional place-and-route iterations, thereby substantially prolonging the design cycle~\cite{kahng2018new}.

Prior work on logic synthesis flow optimization falls into two paradigms, as shown in the upper right of \Cref{fig:comparison}.
The first treats synthesis tuning as combinatorial search over AIG-level operator sequences on ABC~\cite{brayton2010abc}, with methods spanning manual tuning, reinforcement learning (RL)~\cite{hosny2020drills}, Bayesian optimization (BO)~\cite{grosnit2022boils,pei2023alphasyn}, Monte Carlo Tree Search (MCTS)~\cite{yu2020flowtune}, and bandits~\cite{liu2024cbtune}.
Although effective, these methods are confined to fixed action spaces, rely on black-box rewards, and provide limited interpretability regarding how their decisions relate to the current circuit state. These limitations become more pronounced on commercial tools with far richer command vocabularies.
The second paradigm emerges from the growing use of large language models (LLMs) in EDA. Works such as ChatEDA~\cite{wu2024chateda}, ChipNeMo~\cite{liu2023chipnemo}, and ChatLS~\cite{zheng2025chatls} generate complete Tcl scripts or command sequences for commercial tools directly from user intent and tool documentation, thereby avoiding a fixed action space.
However, these end-to-end methods rely on ``one-shot'' generation and produce complete static scripts upfront. As a result, later commands cannot adapt to evolving circuit states after each command, potentially leading to suboptimal results.

A key observation is that commercial synthesis tools support a continuous interaction model: the tool session remains open across commands, allowing an agent to issue commands and observe updated PPA reports after each iteration, making synthesis a natural fit for a closed-loop LLM agent.
This suggests that rather than generating a script upfront or searching over a fixed action space, the agent should interact directly with the tool, diagnosing the current circuit state and retrieving relevant knowledge at each iteration to guide its next decision.
However, two challenges arise:
(1) at each iteration, the LLM must locate relevant content within extensive command documentation, where irrelevant information can cause reasoning errors;
(2) pure LLM reasoning lacks systematic reuse of prior experience, making it difficult to reuse commands proven effective in previous iterations and leading to inefficient long-horizon exploration.

To address these challenges, we present \textbf{SynAct}, an adaptive closed-loop LLM reasoning--acting agent that iteratively optimizes PPA on a commercial synthesis tool, with particular emphasis on WNS.
At each iteration, SynAct conditions its decision on the latest timing reports and diagnostic outputs, mitigating information overload through a multi-layer GraphRAG module~\cite{pu2024customized} that retrieves scenario-relevant command documentation on demand.
To improve exploration efficiency, Bayesian optimization over a GrammarVAE latent space leverages historical optimization experience accumulated in earlier iterations.
Built on ReAct~\cite{yao2022react} for coupled reasoning and action and AutoGen-style multi-agent collaboration~\cite{wu2024autogen} for role coordination, SynAct continuously interprets circuit feedback and produces each optimization command with an explicit rationale, as illustrated in \Cref{fig:comparison}.
Our main contributions are as follows:
\begin{enumerate}
    \item A closed-loop LLM agent that iteratively optimizes PPA by adapting decisions to the current circuit state.
    \item A multi-layer GraphRAG module for scenario-relevant knowledge retrieval, mitigating information overload.
    \item A BO-guided experience refinement mechanism over a GrammarVAE latent space for experience-guided exploration.
    \item SynAct reduces average WNS to 27\% of that from bootstrap synthesis while maintaining balanced area and power trade-offs.
\end{enumerate}

The remainder of this paper is organized as follows. \Cref{sec:prelim} reviews the necessary background. \Cref{sec:framework} presents the SynAct framework, and \Cref{sec:guidance} details its knowledge- and experience-guidance modules. \Cref{sec:running_example} provides a practical example of the framework in operation. \Cref{sec:experiment} reports the experimental results, and \Cref{sec:conclusion} concludes the paper.

%% file: doc/prelim.tex
\section{Preliminaries}
\label{sec:prelim}

\subsection{Logic Synthesis}

Logic optimization and technology mapping are key stages in logic synthesis, respectively restructuring Boolean networks and mapping them to library cells under timing, area, and power constraints. Numerous algorithms have been proposed for these tasks~\cite{brayton2010abc}.
Commercial synthesis tools offer a much richer command space that includes combinational and sequential optimization, retiming, power-aware optimization, and physically aware synthesis~\cite{dc_guide}.
These capabilities create a larger and more context-dependent action space that must be explored effectively to achieve high-quality PPA results.

Because each command alters the circuit state and subsequent actions rely on the updated netlist and PPA reports, synthesis optimization can be formulated as a Markov Decision Process (MDP) $\mathcal{M} = (\mathcal{X}, \mathcal{A}, \mathcal{F}, \mathcal{R})$, forming the basis of our agent design.
\begin{itemize}
    \item $\mathcal{X}$ is the state space, where each state $x_t$ includes the current netlist, PPA summary, and detailed timing, area, and power reports.
    \item $\mathcal{A}$ is the action space of executable synthesis commands.
    \item $\mathcal{F}: \mathcal{X} \times \mathcal{A} \rightarrow \mathcal{X}$ is the transition function realized by the synthesis tool under fixed settings.
    \item $\mathcal{R}: \mathcal{X} \times \mathcal{A} \rightarrow \mathbb{R}$ is the reward function, which uses a weighted aggregation of WNS and other PPA metrics to capture timing, area, and power trade-offs.
\end{itemize}

Bootstrap synthesis is the initialization flow that configures the technology library, reads and elaborates RTL, applies timing constraints, and runs the initial optimization. It excludes recipe-level tuning and yields the common initial state $x_0$ and reference PPA for all compared methods on each design.
Starting from $x_0$, the agent iteratively selects one command $a_t$, executes it within the tool session, and observes the resulting reports as the next state $x_{t+1}$ until the optimization target is met or the budget is exhausted.

\subsection{LLMs for EDA}

LLM-based EDA increasingly combines knowledge retrieval with agentic tool interaction. Retrieval-Augmented Generation (RAG)~\cite{lewis2020retrieval} grounds generation in retrieved knowledge, and its retrieval models can be fine-tuned on EDA tool documentation to improve domain-specific retrieval~\cite{pu2024customized}. Beyond retrieval, ReAct~\cite{yao2022react} interleaves reasoning and action so that observations inform later decisions. AutoGen~\cite{wu2024autogen} supports coordination among specialized agents through structured conversations in complex workflows.
In practice, ChatEDA~\cite{wu2024chateda} decomposes user requests before generating and executing RTL-to-GDSII scripts, while ChipNeMo~\cite{liu2023chipnemo} combines domain adaptation and retrieval for tasks including EDA script generation. ChatLS~\cite{zheng2025chatls} uses retrieved tool knowledge and chain-of-thought prompting to customize synthesis scripts, whereas LLSM~\cite{huang2025llsm} uses EDA-guided prompting to integrate circuit information extracted from RTL with structural features. These studies show that LLMs can retrieve domain knowledge, reason about design tasks, coordinate specialized agents, and invoke EDA tools, laying the foundation for SynAct's closed-loop PPA optimization.

%% file: doc/algo.tex
\section{SynAct Framework}
\label{sec:framework}
\begin{figure*}[tb!]
    \centering
    \includegraphics[width=\textwidth]{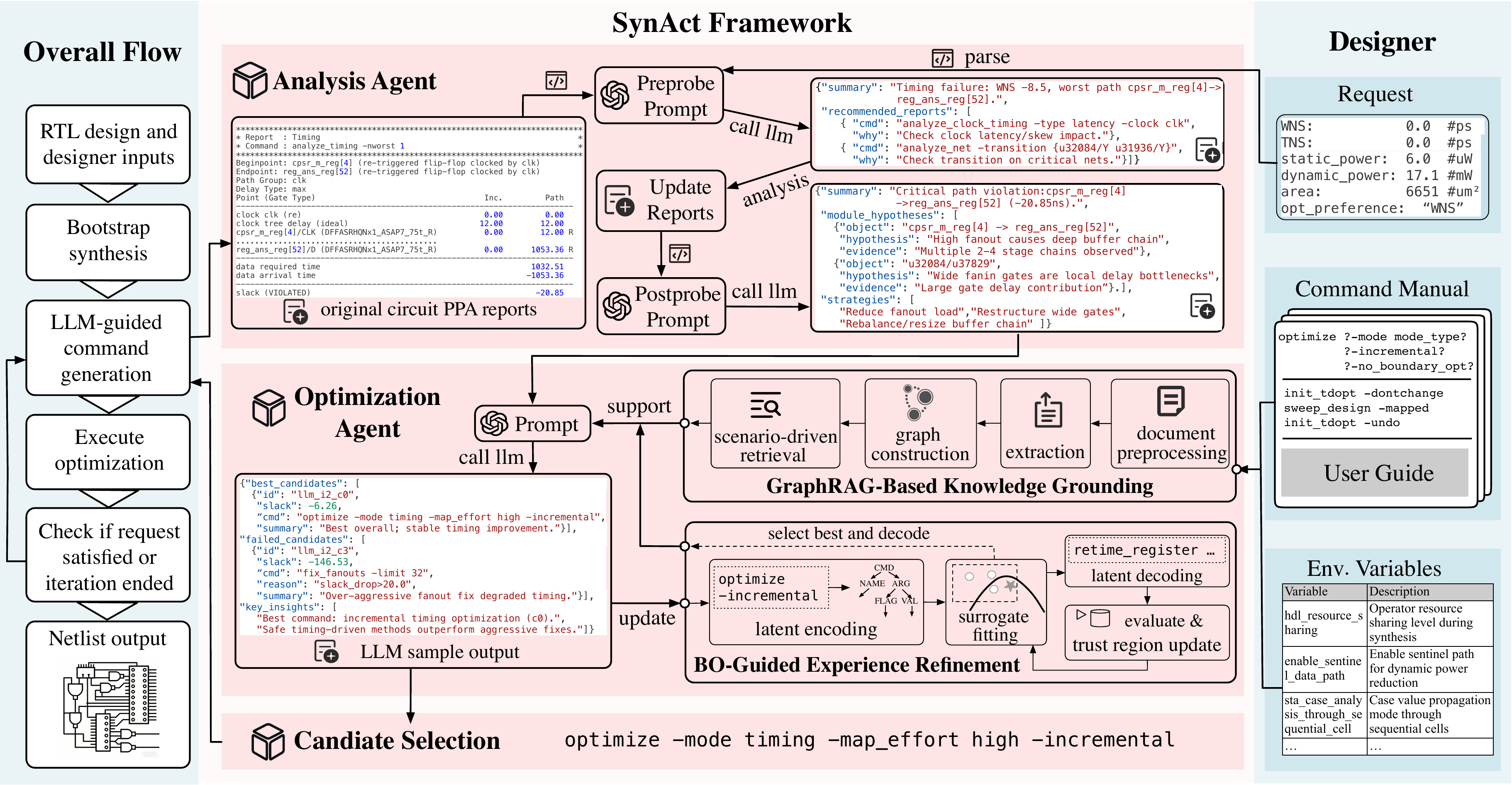}
    \caption{The overall framework of SynAct.}

    \label{fig:framework}
\end{figure*}

\subsection{Overview}
The left side of \Cref{fig:framework} shows the overall flow. Given an RTL design and designer inputs, SynAct first performs bootstrap synthesis to establish the initial circuit state. It then iterates through LLM-guided command generation, optimization execution, and termination checking until the request is satisfied or the iteration budget $n_{\mathrm{iter}}$ is exhausted. The resulting netlist is then written out.

The right side lists the designer inputs, including a request, the command manual, and environment variables. The request specifies the optimization preferences and quantitative PPA targets, while the manual and variables provide tool-specific information for command generation. SynAct can prioritize any selected PPA metric. In this work, we use \textbf{WNS} as the primary objective and treat the remaining metrics as secondary objectives.

The center of \Cref{fig:framework} details three main components of SynAct. Inspired by ReAct~\cite{yao2022react}, the Analysis Agent examines the current reports and produces a diagnosis. The Optimization Agent combines this diagnosis with retrieved tool knowledge and historical experience to generate command candidates. Candidate Selection chooses the best safe command from the evaluated candidates for execution. Executing the command updates the circuit state, and the resulting reports inform the next iteration, forming a closed reasoning--acting loop. The following subsections detail these three components.

\subsection{Analysis Agent}
Existing search-based methods guide optimization primarily using scalar PPA summaries~\cite{grosnit2022boils,pei2023alphasyn,liu2024cbtune}, but these summaries provide limited information, making it difficult to uncover the circuit-level causes of PPA violations. SynAct therefore introduces an Analysis Agent that inspects detailed timing, area, and power reports to diagnose potential performance bottlenecks through two stages: Preprobe and Postprobe.

In Preprobe, the agent parses raw report logs of the current design to identify performance bottlenecks. 
When the available data lacks sufficient detail, it proactively constructs follow-up probes using \texttt{analyze\_*} commands described in the tool documentation to obtain the necessary information.
For precise and context-aware diagnosis, the agent is guided to generate targeted commands for specific design objects such as clock domains, timing constraints, circuit instances, or buffer chains through the following prompt:
\begin{promptbox}\small
    \textbf{System:} role = analysis agent (PREPROBE); diagnose root causes and propose \texttt{analyze\_*} probes; prefer object-level Tcl-braced probes\\
    \textbf{User:} user request, current/previous metrics, raw report log, and preprobe policy\\
    \textbf{Expected JSON:} \{~summary, recommended\_probes~\}
\end{promptbox}

A typical output may include probes such as:
\begin{quote}
\small\ttfamily
analyze\_clock\_timing -type latency\par
analyze\_inst -pin \{...\}\par
analyze\_buffer\_chain -from \{...\}
\end{quote}

In Postprobe, the outputs collected by executing the probes from the previous stage are incorporated into the user prompt to help the agent refine its diagnosis.
The refined result is then returned as a structured diagnosis that forms part of the Optimization Agent's input.
For example, it may conclude that one path group dominates the timing bottleneck and that the next actions should focus on buffer optimization, cell sizing, or local restructuring under area and power guardrails.
The corresponding Postprobe prompt is:
\begin{promptbox}\small
    \textbf{System:} role = analysis agent (POSTPROBE); refine diagnosis using collected probe evidence\\
    \textbf{User:} user request, raw report text, extra analysis reports\\
    \textbf{Expected JSON:} \{~summary, goals, violations, module\_hypotheses, suggested\_strategies, reflection~\}
\end{promptbox}

\subsection{Optimization Agent}
The Optimization Agent takes the structured diagnosis and user request as input and generates executable synthesis command candidates.
This constitutes our second key design: rather than optimizing over a small fixed action set, SynAct performs context-aware command generation within a large, structured command space. 
This generation process is supported by two complementary guidance modules, which are detailed in \Cref{sec:guidance}:
\begin{itemize}
    \setlength{\itemsep}{0pt}
    \item \textbf{RAG-grounded candidates} $\mathcal{C}_{\mathrm{RAG}}$, which are derived from scenario-relevant manual sections retrieved by GraphRAG;
    \item \textbf{BO-seeded candidates} $\mathcal{C}_{\mathrm{BO}}$, which adapt historically effective command patterns with minimal syntax changes.
\end{itemize}
By combining these sources, the agent generates commands that reflect the current diagnosis while leveraging retrieved knowledge and prior experience. 
Each candidate is accompanied by a concise rationale. 
The corresponding structured prompt is:
\begin{promptbox}\small
    \textbf{System:} role = optimization agent; generate command candidates\\
    \textbf{User:} user request, Postprobe diagnosis, RAG-retrieved manual sections, BO seed recommendation\\
    \textbf{Expected JSON:} \{~recommended\_candidates, rationale~\}
\end{promptbox}

\subsection{Candidate Selection}\label{sec:candidate_selection} 
Given the candidates produced by the Optimization Agent, SynAct first performs safety filtering, removing those whose WNS falls beyond a preset threshold (e.g., $-20$ ps) or whose reward drops below $0.2 \times r_{\text{boot}}$, where $r_{\text{boot}}$ is the initial reward from bootstrap synthesis. 
Among the remaining safe candidates, it then selects the one with the highest score (defined below). 
If no candidate passes filtering, the previous checkpoint is retained.

Each candidate command $C$ is evaluated by the score
\begin{equation}\label{eq:score}
    \mathrm{score}(C) = r(C) + \kappa\sigma(z) - \rho(C),
\end{equation}
where $r(C)$ is the reward measuring objective satisfaction, $\kappa$ controls exploration strength, and $\rho(C)$ penalizes recent repetitions (waived for large WNS improvements). 
Here $\sigma(z)$ represents a local uncertainty estimate from the BO surrogate at latent point $z$, encouraging exploration of under-explored regions. 
All candidates, whether from $\mathcal{C}_{\mathrm{BO}}$ or $\mathcal{C}_{\mathrm{RAG}}$, are encoded into this shared latent space, enabling the surrogate to generalize learned experience and provide consistent exploration guidance across both sources.
The latent space and surrogate model are detailed in \Cref{sec:bo_guided_experience_refinement}.
This score balances high reward, moderate exploration, and redundancy avoidance.

The reward $r\in(0,1]$ aggregates violations across all objectives:
\begin{align}
    r \label{eq:reward} & = \exp\!\left(-\sum_i w_i \log(1+\mathrm{err}_i)\right), \\[4pt]
    \mathrm{err}_i      & =
    \begin{cases}
        \max(0,\, \mathrm{target}_i - \mathrm{metric}_i),                              & i \in \{\text{WNS}, \text{TNS}\},                       \\[3pt]
        \dfrac{\max(0,\, \mathrm{metric}_i - \mathrm{target}_i)}{|\mathrm{target}_i|}, & i \in \{\text{area}, P_{\text{dyn}}, P_{\text{stat}}\},
    \end{cases}
\end{align}
where $\mathrm{target}_i$ is the user-specified target and $\mathrm{metric}_i$ is the current measured value.
Timing metrics (WNS, TNS) are maximized, while area and power are minimized and normalized by target magnitude.
The $\log(1{+}\mathrm{err}_i)$ term compresses large violations to reduce their influence in the aggregate reward.
Each weight $w_i$ is assigned a base value, upweighted for the user-selected primary objective, and then normalized to ensure $\sum_i w_i = 1$.
Thus $r=1$ when all targets are met and decreases faster for larger or higher-priority violations.

\section{Knowledge- and Experience-Guided Optimization}
\label{sec:guidance}

The Optimization Agent is supported by two complementary modules shown in \Cref{fig:framework}: GraphRAG grounds candidate generation in scenario-relevant tool knowledge, while BO reuses historical synthesis experience in a GrammarVAE latent space. This section details these two modules.

\subsection{GraphRAG-Based Knowledge Grounding}
The Optimization Agent requires precise, context-aware domain knowledge to generate effective synthesis commands.
A natural approach is to retrieve relevant documentation via RAG; however, forming effective synthesis strategies requires connecting applicable scenarios, specific commands, and configurable
variables scattered across disparate documentation sections. 
This structured, multi-hop reasoning process goes beyond the ability of traditional vector-similarity retrieval.
To address this, we develop a multi-layer \textbf{GraphRAG module} that unifies graph-based reasoning and retrieval augmentation~\cite{edge2024local,wu2025medical}, detailed in the following four parts.

\begin{figure}[tb!]
    \centering
    \includegraphics[width=0.48\textwidth]{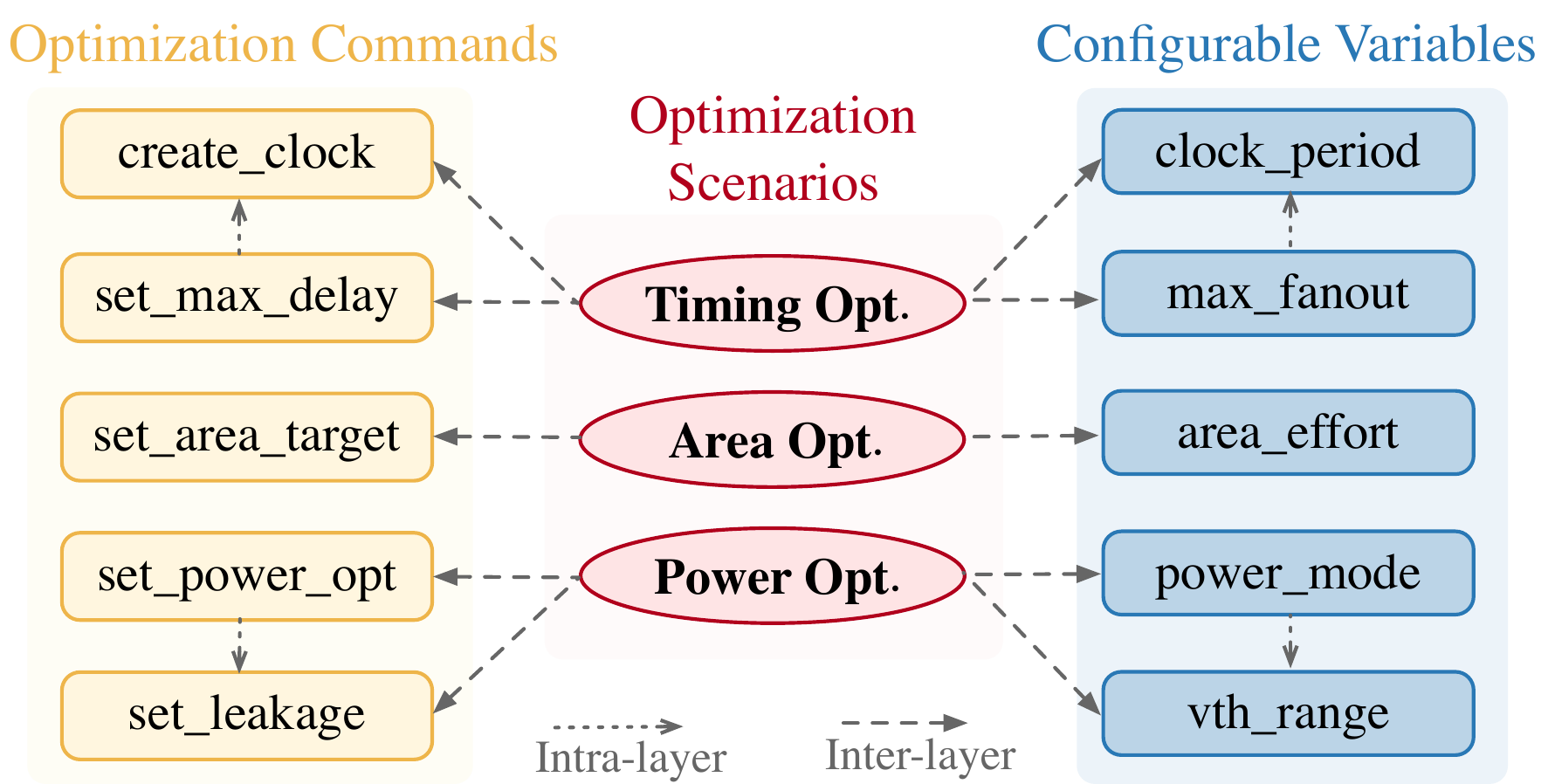}

    \caption{
        Three-layer structure of the proposed GraphRAG framework.
        The optimization scenario layer (center) connects related commands (left) and configurable variables (right), forming inter- and intra-layer relationships.
    }
    \label{fig:graph_structure}
\end{figure}

\minisection{Document Preprocessing}
Building the GraphRAG module begins with extracting relevant content from the synthesis tool documentation.
We manually curate document sections related to optimization commands and categorize them into three functional roles:
\begin{enumerate}
    \item Applicable scenarios: describe design goals (e.g., timing, area, power) with typical associated commands and variables;
    \item Executable commands: define specific tool operations, including syntax, structure, and invocation patterns;
    \item Configurable variables: specify tunable parameters, valid ranges, and default configurations.
\end{enumerate}

This taxonomy follows practical EDA usage: optimizations are scenario-driven, executed through commands, and tuned by variables.
Based on this categorization, we organize the extracted content into a three-layer graph of scenarios, commands, and variables.

\minisection{Entity and Relationship Extraction}
An LLM is prompted to convert the unstructured text into structured entities belonging to the three layers defined above.
We introduce three disjoint sets of entities corresponding to the three layers,
where $\mathcal{E}_S = \{s_1, s_2, \ldots, s_{n_S}\}$, $\mathcal{E}_C = \{c_1, c_2, \ldots, c_{n_C}\}$, and $\mathcal{E}_V = \{v_1, v_2, \ldots, v_{n_V}\}$ represent scenarios, commands, and variables, respectively.

The semantic structure of the knowledge graph is built through two complementary processes: intra-layer and inter-layer relationship identification.
For intra-layer relationships, we use an LLM-based identification function to detect semantic connections within a single layer, defined as follows:
\begin{gather}
    \mathcal{R}^{(intra)} = \mathcal{R}_{CC} \cup \mathcal{R}_{VV}, \\
    \mathcal{R}_{CC} = \{(c_i, c_j, r_{ij}) \mid c_i, c_j \in \mathcal{E}_C,\, \text{LLM}_{rel}(c_i, c_j) = \text{True}\}, \\
    \mathcal{R}_{VV} = \{(v_i, v_j, r'_{ij}) \mid v_i, v_j \in \mathcal{E}_V,\, \text{LLM}_{rel}(v_i, v_j) = \text{True}\}.
\end{gather}
Here, $\text{LLM}_{rel}(\cdot,\cdot)$ determines whether two entities share a semantic dependency based on their descriptions and usage patterns.

For inter-layer relationships, scenarios are linked to related commands and variables that co-occur in documentation examples, formalized as follows:
\begin{gather}
    \mathcal{R}^{(inter)} = \mathcal{R}_{SC} \cup \mathcal{R}_{SV}, \\
    \mathcal{R}_{SC} = \{(s, c) \mid s \in \mathcal{E}_S,\, c \in \mathcal{E}_C,\, c \in \mathcal{C}_s^{(cmds)}\}, \\
    \mathcal{R}_{SV} = \{(s, v) \mid s \in \mathcal{E}_S,\, v \in \mathcal{E}_V,\, v \in \mathcal{C}_s^{(vars)}\}.
\end{gather}
A link $(s,c)$ or $(s,v)$ is established when the LLM identifies co-occurrence between scenario $s$ and the corresponding command or variable in application examples.

\minisection{Hierarchical Knowledge Graph Construction}
Based on the extracted entities and relations, we construct a hierarchical knowledge graph defined as:
\begin{align}
    \mathcal{G} & = (\mathcal{E}, \mathcal{R}),
\end{align}
where $\mathcal{E} = \mathcal{E}_S \cup \mathcal{E}_C \cup \mathcal{E}_V$ denotes the set of all entities, which is partitioned into disjoint subsets of scenarios, commands, and variables, and $\mathcal{R} = \mathcal{R}^{(\text{intra})} \cup \mathcal{R}^{(\text{inter})}$ represents intra-layer and inter-layer relations. The scenario layer acts as the central hub connecting the command and variable layers, providing the foundation for subsequent reasoning and retrieval, as illustrated in~\Cref{fig:graph_structure}.

\minisection{Scenario-Driven Retrieval}
Before retrieval, the \textbf{Analysis Agent} diagnoses optimization bottlenecks and generates an analysis report $\mathcal{A}$, which serves as the input query. Given $\mathcal{A}$, GraphRAG retrieves relevant synthesis knowledge via four compact steps:
\begin{enumerate}
    \item Scenario description generation: An LLM summarizes an optimization scenario description $d_{\mathrm{scene}}$ from $\mathcal{A}$.
    \item Semantic embedding: scenario description $d_{\mathrm{scene}}$ and scenario entity $s$ are encoded into vectors using a domain-customized text embedding model~\cite{pu2024customized}, fine-tuned on EDA tool documentation via supervised contrastive learning. This model better captures EDA terminology and semantics than general-purpose embeddings, ensuring all entities/descriptions in GraphRAG share a consistent EDA-aware vector space for precise retrieval.
    \item Scenario selection: Top-$k$ relevant scenarios are retrieved by semantic similarity:
          \begin{equation}
              \mathcal{S}^* = \underset{\mathcal{S} \subseteq \mathcal{E}_S, |\mathcal{S}|=k}{\arg\max} \sum_{s \in \mathcal{S}} \text{sim}(\text{Embed}(d_{\mathrm{scene}}), \text{Embed}(s)).
          \end{equation}
    \item Knowledge expansion: Retrieve commands and variables directly connected to $\mathcal{S}^*$ as the initial associated set $\mathcal{C}_0$/$\mathcal{V}_0$, and further retrieve their top-$k_{\mathrm{intra}}$ intra-layer neighbors via k-nearest neighbor (KNN) to form the final expanded set. Specifically:

          {\footnotesize
          \begin{align}
              \mathcal{C}_0 = \{c \mid (s, c) \in \mathcal{R}_{SC},\, s \in \mathcal{S}^*\},
              \; \mathcal{C}^* = \mathcal{C}_0 \cup \text{KNN}(\mathcal{C}_0, k_{\mathrm{intra}}), \\
              \mathcal{V}_0 = \{v \mid (s, v) \in \mathcal{R}_{SV},\, s \in \mathcal{S}^*\},
              \; \mathcal{V}^* = \mathcal{V}_0 \cup \text{KNN}(\mathcal{V}_0, k_{\mathrm{intra}}).
          \end{align}}
          Here, $k_{\mathrm{intra}}$ denotes the number of intra-layer neighbors retrieved for each seed entity.
\end{enumerate}
The final retrieved knowledge set is $\mathcal{K} = \mathcal{S}^* \cup \mathcal{C}^* \cup \mathcal{V}^*$, providing the Optimization Agent with concise, scenario-specific, non-redundant EDA synthesis knowledge.

The hierarchical GraphRAG mechanism tightly couples knowledge grounding with design diagnosis, which allows the Optimization Agent to leverage structured, reliable synthesis knowledge for improved reasoning accuracy and decision reliability.

\subsection{BO-Guided Experience Refinement}\label{sec:bo_guided_experience_refinement}

\begin{figure}[tb!]
    \centering
    \includegraphics[width=0.48\textwidth]{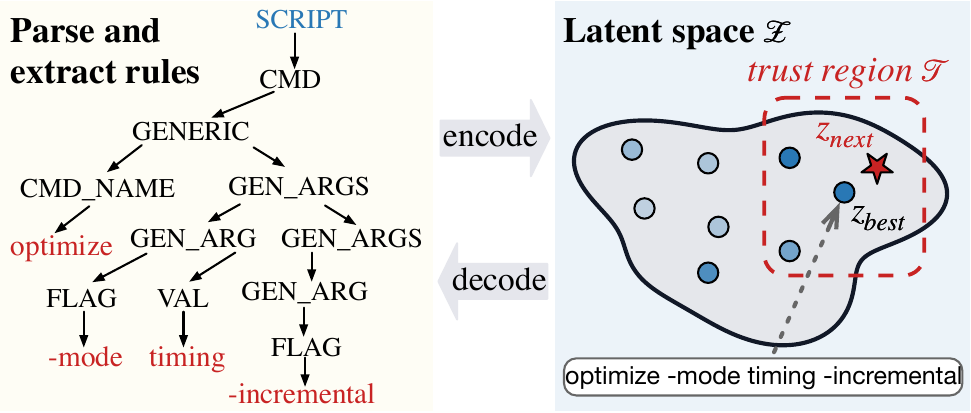}
    \caption{GrammarVAE encoding and latent-space BO.
        A command is parsed into a grammar tree and encoded into latent space $\mathcal{Z}$;
        BO then identifies $z_{\mathrm{best}}$ and proposes $z_{\mathrm{next}}$ within the trust region $\mathcal{T}$,
        which is decoded back into a command recommendation for the Optimization Agent.}
    \label{fig:grammar-vae}
\end{figure}

\begin{algorithm}[tb!]
    \caption{BO-Guided Experience Refinement}
    \label{alg:bo_refinement}
    \begin{algorithmic}[1]

        \Require Experience log $\mathcal{D}$, trust region $\mathcal{T}$, last commit reward $r_{\mathrm{prev}}$, diagnosis $\Phi$, RAG docs $\mathcal{R}$

        \Ensure Updated log $\mathcal{D}'$, updated trust region $\mathcal{T}'$, BO seed $\hat{C}_{\mathrm{BO}}$

        \State /* \textbf{Surrogate Fitting} */
        \State $w_i \leftarrow \min(w_{\max},\, 1 + \lambda\,\max(0, r_i))$ \textbf{for each} $(z_i, r_i) \in \mathcal{D}$\label{algline:bo_w}
        \State Refit RBF surrogate on $\mathcal{D}$ to obtain $\mu(z)$, $\sigma(z)$\label{algline:bo_refit}

        \State /* \textbf{UCB Acquisition and Latent Decoding} */
        \State Sample pool $\mathcal{P} \subset \mathcal{T}$\label{algline:bo_pool}
        \State $z_{\mathrm{next}} \leftarrow \arg\max_{z \in \mathcal{P}}\,[\mu(z) + \kappa_{\mathrm{acq}}\,\sigma(z)]$\label{algline:bo_ucb}
        \State $\hat{C}_{\mathrm{BO}} \leftarrow \mathrm{Dec}(z_{\mathrm{next}})$ \Comment{GrammarVAE decode}\label{algline:bo_decode}
        \State $\mathcal{C} \leftarrow \mathrm{OptAgent}(\hat{C}_{\mathrm{BO}},\, \Phi,\, \mathcal{R})$\label{algline:bo_optagent}

        \State /* \textbf{Evaluation and Trust Region Update} */
        \State $\mathcal{D}' \leftarrow \mathcal{D}$
        \For{$C \in \mathcal{C}$}\label{algline:bo_eval}
        \State Execute $C$; record $r(C)$\label{algline:bo_exec}

        \State $z \leftarrow \mathrm{Enc}(\mathrm{Norm}(C))$ \Comment{GrammarVAE encode}\label{algline:bo_reencode}
        \State $\mathcal{D}' \leftarrow \mathcal{D}' \oplus (z, r(C))$ \Comment{$\oplus$: insert or replace}\label{algline:bo_log}
        \EndFor
        \State $\mathcal{C}_{\mathrm{safe}} \leftarrow \{C \in \mathcal{C} \mid C \text{ passes safety filter}\}$ \Comment{\Cref{sec:candidate_selection}}\label{algline:bo_safe}
        \State $C^\star \leftarrow \arg\max_{C \in \mathcal{C}_{\mathrm{safe}}}\,\mathrm{score}(C)$ \Comment{see \Cref{eq:score}}\label{algline:bo_commit}

        \State $z_{\mathrm{best}} \leftarrow \mathrm{Enc}(\mathrm{Norm}(C^\star))$\label{algline:bo_zbest}
        \State Recenter $\mathcal{T}'$ on $z_{\mathrm{best}}$; expand if $r(C^\star) > r_{\mathrm{prev}}$, else shrink\label{algline:bo_trust}

        \State \Return $(\mathcal{D}',\, \mathcal{T}',\, \hat{C}_{\mathrm{BO}})$\label{algline:bo_return}

    \end{algorithmic}
\end{algorithm}

Commands validated in earlier iterations provide design-specific evidence for promising optimization directions~\cite{grosnit2022boils,pei2023alphasyn,liu2024cbtune}. Inspired by this, the Optimization Agent reuses executed commands and their feedback as experience. In practice, reapplying a previously effective command can yield further gains and sometimes outperform one newly generated from retrieved documentation. This motivates an experience-guided search based on Bayesian optimization (BO), which uses accumulated evaluations to efficiently identify promising commands.

Applying BO directly to discrete command strings remains inefficient because the search space is large and similar commands can exhibit different optimization behavior. Inspired by grammar-based VAEs that map discrete grammatical structures into continuous spaces~\cite{kusner2017grammar,lynch2020program}, we adapt GrammarVAE to encode synthesis commands for BO. This representation enables smooth optimization while preserving syntactic validity~\cite{dai2018syntax}. BO then iteratively fits a surrogate model, proposes and evaluates candidates, and updates its data to focus on empirically effective regions.

\minisection{Latent Encoding}
Each synthesis command $C$ is normalized by $\mathrm{Norm}(\cdot)$ and encoded into a latent vector $z = \mathrm{Enc}(\mathrm{Norm}(C)) \in \mathbb{R}^{d}$ via the GrammarVAE encoder~\cite{kusner2017grammar}, where $z$ is a point in latent space $\mathcal{Z} \subseteq \mathbb{R}^d$.
As shown in \Cref{fig:grammar-vae}, a command is first parsed into a context-free grammar tree, then compressed into $z$, such that nearby points in $\mathcal{Z}$ correspond to grammatically similar commands.
Because synthesis-command syntax encodes the optimization mode, transformation scope, effort level, and parameter configuration, these grammar-preserving neighborhoods provide a structured local prior that BO calibrates with measured synthesis rewards.
GrammarVAE is pre-trained offline on a corpus of synthesis commands using the standard VAE objective (reconstruction and KL regularization); during BO, each command is mapped to the encoder's posterior mean for a consistent latent representation.

\begin{mydefinition}[Trust Region $\mathcal{T}$]
BO limits its search to an axis-aligned trust region $\mathcal{T} \subseteq \mathcal{Z}$, centered at the latent point of the current best command $z_{\mathrm{best}}$, to focus optimization on the most promising area.
The radius of $\mathcal{T}$ adapts across iterations: it expands when performance improves and shrinks otherwise.
\end{mydefinition}

\begin{mydefinition}[Experience Log $\mathcal{D}$]
The experience log $\mathcal{D} = \{(z_i, r_i)\}_{i=1}^{N}$ stores the latent encoding and reward of each executed command $C_i$, with $z_i = \mathrm{Enc}(\mathrm{Norm}(C_i))$ and $r_i$ defined in \Cref{eq:reward}.
During bootstrap synthesis, $\mathcal{D}$ is initialized from the default \texttt{optimize} command, and $\mathcal{T}$ is centered at its latent point.
\end{mydefinition}


\Cref{alg:bo_refinement} illustrates how the refinement procedure is executed within one iteration: it takes $\mathcal{D}$, $\mathcal{T}$, last commit reward $r_{\mathrm{prev}}$, the Analysis Agent's diagnosis $\Phi$, and RAG-retrieved documents $\mathcal{R}$ as input, and returns updated $\mathcal{D}'$, $\mathcal{T}'$, and BO seed $\hat{C}_{\mathrm{BO}}$.

\minisection{Surrogate Fitting}
At each iteration, a reward-weighted RBF surrogate~\cite{gutmann2001radial} is refit from scratch on the full $\mathcal{D}$ (lines~\ref{algline:bo_w}--\ref{algline:bo_refit}).
Each observation $(z_i, r_i)$ is assigned a weight $w_i = \min(w_{\max}, 1 + \lambda\,\max(0, r_i))$, where $\lambda \ge 0$ scales how aggressively positive rewards upweight a point in the fit and $w_{\max}$ caps each $w_i$, so that high-reward regions receive greater influence while no single sample dominates.
The surrogate estimates the expected reward and local uncertainty through kernel-weighted averages:
\begin{equation}
    \mu(z) = \frac{\sum_i w_i K(z,z_i) r_i}{\sum_i w_i K(z,z_i)},
    ~
    \sigma^2(z) = \frac{\sum_i w_i K(z,z_i) [r_i - \mu(z)]^2}{\sum_i w_i K(z,z_i)},
\end{equation}
where $K(z,z_i)=\exp(-\|z-z_i\|^2/2\ell^2)$ is the RBF kernel.
The resulting $\sigma(z)$ is smoothed and clipped for numerical stability and serves as the local uncertainty estimate used by the commitment score described in \Cref{sec:candidate_selection}.

\minisection{UCB Acquisition and Latent Decoding}
BO samples a pool $\mathcal{P} \subset \mathcal{T}$ and selects the next latent point by maximizing the UCB acquisition function~\cite{srinivas2010gaussian} (lines~\ref{algline:bo_pool}--\ref{algline:bo_ucb}):
\begin{equation}
    \alpha(z) = \mu(z) + \kappa_{\mathrm{acq}}\,\sigma(z),
    \label{eq:bo_ucb}
\end{equation}
where $\mu(z)$ exploits known high-reward regions and $\kappa_{\mathrm{acq}}\,\sigma(z)$ promotes exploration of regions with high local uncertainty.
The selected $z_{\mathrm{next}}$ is decoded by GrammarVAE into a command string $\hat{C}_{\mathrm{BO}}$ (line~\ref{algline:bo_decode}) and passed to the Optimization Agent as the BO seed recommendation.
While GrammarVAE decoding enforces grammatical validity, the decoded command may still contain tool-level errors (e.g., unsupported flags or invalid parameter ranges).
The Optimization Agent retains $\hat{C}_{\mathrm{BO}}$ as an explicit conditioning input and refines it with retrieved tool documentation to produce $\mathcal{C}_{\mathrm{BO}}$, while independently generating $\mathcal{C}_{\mathrm{RAG}}$ from the retrieved knowledge; together they form the full candidate set $\mathcal{C} = \mathcal{C}_{\mathrm{BO}} \cup \mathcal{C}_{\mathrm{RAG}}$ (line~\ref{algline:bo_optagent}).

\minisection{Evaluation and Trust Region Update}
All candidates in $\mathcal{C}$ are executed from the same parent checkpoint and their rewards recorded (lines~\ref{algline:bo_eval}--\ref{algline:bo_exec}).
Each candidate is then re-encoded and logged into $\mathcal{D}'$ (lines~\ref{algline:bo_reencode}--\ref{algline:bo_log}), including those that fail or yield poor results, as broader latent-space coverage helps BO distinguish effective from ineffective regions and improve future proposals.
The committed command $C^\star$ is selected by first filtering $\mathcal{C}$ for safe candidates $\mathcal{C}_{\mathrm{safe}}$ (line~\ref{algline:bo_safe}), and then maximizing the score function $\mathrm{score}(C)$ defined in \Cref{eq:score} (line~\ref{algline:bo_commit}).
The trust region $\mathcal{T}'$ is recentered on $z_{\mathrm{best}} = \mathrm{Enc}(\mathrm{Norm}(C^\star))$ and expanded if $r(C^\star) > r_{\mathrm{prev}}$, otherwise shrunk (lines~\ref{algline:bo_zbest}--\ref{algline:bo_trust}).


\section{Running Example}\label{sec:running_example}

\begin{figure}[tb!]
    \centering
    \includegraphics[width=0.48\textwidth]{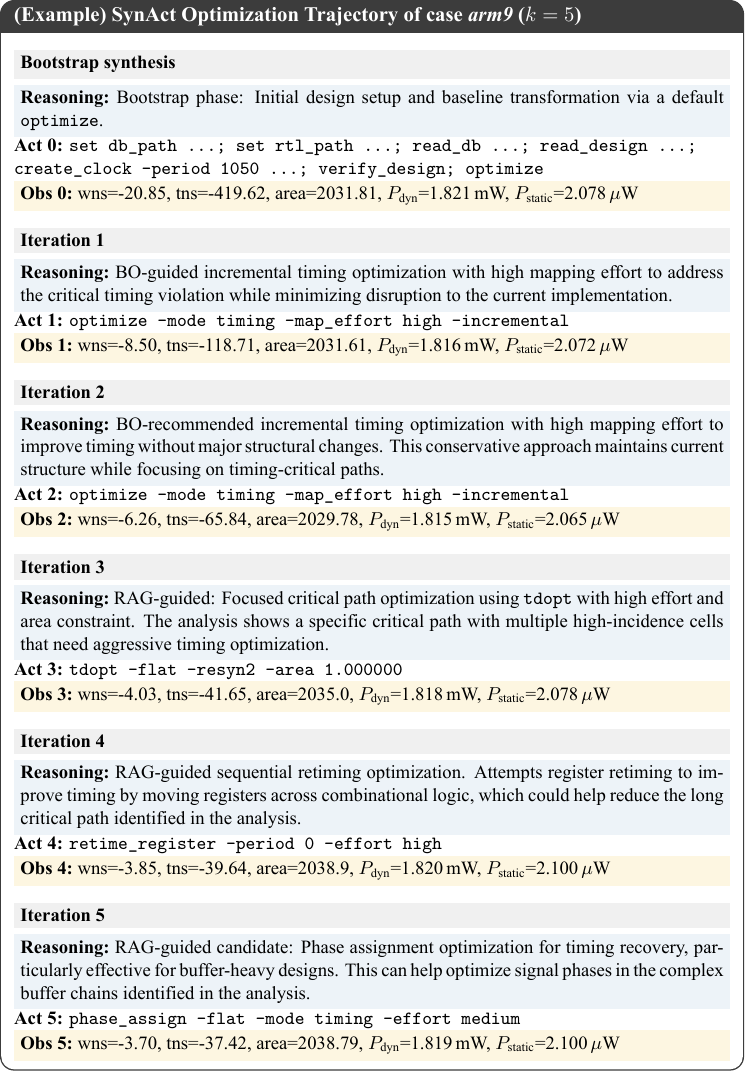}
    \caption{An actual SynAct run on \textit{arm9}, showing the reasoning, committed action, and tool observation at each iteration.}
    \label{fig:synact_react_tcl}
\end{figure}

To illustrate the practical use of SynAct, \Cref{fig:synact_react_tcl} presents a five-iteration run on \textit{arm9}. Starting from the initial synthesis report, SynAct follows a reasoning, acting, and observation loop. At each iteration, the LLM interprets the tool feedback, explains the optimization intent, and produces an executable Tcl command. After execution, the updated circuit state is reflected in new synthesis reports, which serve as observations for the next iteration.

In the first two iterations, the LLM follows BO guidance and selects incremental timing optimization to limit changes to the current implementation. It then uses updated timing observations and retrieved tool knowledge to select \texttt{tdopt}, register retiming, and phase assignment. The figure records the reasoning, committed command, and PPA observation at each iteration. Over the five iterations, WNS changes from $-20.85$ ps to $-3.70$ ps.

%% file: doc/exp.tex
\section{Experiment}
\label{sec:experiment}

\subsection{Experimental Setup}

All experiments are conducted on a workstation equipped with an Intel\textregistered~Xeon\textregistered~Gold 6226R CPU (2.90\,GHz) and an NVIDIA GeForce RTX 3090 GPU.
SynAct is implemented in Python with PyTorch for neural modules and connects to large language models via an OpenAI-compatible API supporting both commercial and internal endpoints. 
It integrates with \altisyn, the commercial logic synthesis tool from ZeniSyn Design Systems~\cite{zenisyn}, for automated command execution and report collection.
The synthesis flow employs the standard-cell library from the 7~nm ASAP7 edge-cut PDK~\cite{EDA-2016clarkasap7}.

SynAct is implemented as a multi-agent system comprising an Analysis Agent for report-driven diagnosis and an Optimization Agent for command generation.
The Optimization Agent performs Bayesian optimization in a GrammarVAE latent space to efficiently explore discrete synthesis commands and utilizes a Neo4j-based multi-layer GraphRAG module for structured knowledge retrieval.
Semantic retrieval within this module uses the transformer model fine-tuned for EDA documentation in~\cite{pu2024customized}.

For evaluation, we use $14$ open-source RTL designs collected from OpenCores~\cite{opencores}. 
\Cref{tab:Statistics} summarizes the benchmark statistics obtained after importing each design and running a bootstrap synthesis pass, including the number of nets, leaf, sequential, buffer, and inverter cell instances, and the configured clock period (CPD).
Appropriate clock periods are assigned according to design size and fixed across all methods to ensure sufficient optimization margin.
Across all experiments, we set WNS as the primary optimization objective, with quantitative targets specified at the input stage (a configuration example is shown in \Cref{fig:framework}).
Additionally, we track TNS, area, dynamic power, and static power as secondary metrics to assess overall PPA quality.
To account for stochastic variation, each experiment is repeated five times, and the average results are reported unless otherwise specified.

\begin{table}[tb!]
    \footnotesize
    \centering
    \caption{Benchmark statistics.}
    \label{tab:Statistics}
    
    \renewcommand{\arraystretch}{1.1}
    \resizebox{.94\linewidth}{!}
    { 
        \begin{threeparttable}
            \footnotesize
            \begin{tabular}{|c|cccccc|}
              \hline

              Benchmarks & \#Nets & \#Leaf & \#Seq. & \#Buf. & \#Inv. & CPD.\tnote{1} \\ \hline\hline
              \textit{uart16550} & 3057 & 3006 & 605 & 143 & 336 & 310 \\ 
              \textit{picorv32} & 9193 & 9091 & 1556 & 268 & 688 & 390 \\
              \textit{yacc} & 10504 & 10363 & 1102 & 487 & 1285 & 560 \\ 
              \textit{aes\_core} & 11816 & 11557 & 530 & 521 & 1091 & 365 \\               \textit{wb2axip} & 12539 & 11381 & 1900 & 257 & 1221 & 335 \\ 
              \textit{wb\_conmax} & 18170 & 17040 & 770 & 492 & 1354 & 365 \\ 
              \textit{arm9} & 19267 & 19195 & 1222 & 823 & 2634 & 1050 \\ 
              \textit{sha3} & 24927 & 24161 & 2949 & 674 & 2903 & 365 \\ 
              \textit{spimaster} & 38365 & 38313 & 8705 & 862 & 3909 & 320 \\ 
              \textit{ethernet} & 44765 & 44637 & 10543 & 1097 & 3390 & 385 \\ 
              \textit{ecg} & 53613 & 53071 & 7676 & 1116 & 7807 & 340 \\ 
              \textit{linkruncca} & 69138 & 69134 & 16606 & 1720 & 4879 & 480 \\ 
              \textit{xge\_mac} & 81564 & 81358 & 21667 & 29 & 5896 & 135 \\ 
              \textit{fft256} & 198981 & 198897 & 42432 & 3129 & 21095 & 490 \\ \hline
            \end{tabular}
            \begin{tablenotes}
                \item[1] CPD: Clock period (ps).
            \end{tablenotes}
        \end{threeparttable}
    }
\end{table}

\subsection{Overall PPA Improvement}

\begin{table}[tb!]
    \footnotesize
    \centering
    \caption{Original PPA metrics after bootstrap synthesis.}
    \label{tab:original_statistics}
    
    \renewcommand{\arraystretch}{1.1}
    \resizebox{\linewidth}{!}
    { 
        \begin{threeparttable}
            \footnotesize
            \begin{tabular}{|c|ccccc|}
              \hline

              Benchmarks & WNS(ps) & TNS(ps) & Area ($\mu$m$^2$) & DP (mW)\tnote{1} & SP ($\mu$W)\tnote{2}\\ \hline\hline
              \textit{uart16550} & -7.45 & -10.10 & 400.94 & 1.572 & 0.392 \\ 
              \textit{picorv32} & -15.70 & -3132.97 & 1128.68 & 3.529 & 1.248 \\ 
              \textit{yacc} & -26.86 & -1592.03 & 1153.77 & 4.095 & 1.300 \\ 
              \textit{aes\_core} & -8.03 & -482.04 & 1192.06 & 3.695 & 1.316 \\ 
              \textit{wb2axip} & -20.54 & -12344.79 & 1325.09 & 5.158 & 1.169 \\ 
              \textit{wb\_conmax} & -11.35 & -3840.85 & 1698.47 & 4.399 & 1.481 \\ 
              \textit{arm9} & -20.85 & -419.62 & 2031.81 & 1.821 & 2.078 \\ 
              \textit{sha3} & -14.41 & -3446.74 & 2892.50 & 17.930 & 2.874 \\ 
              \textit{spimaster} & -16.96 & -151.50 & 5459.98 & 11.290 & 6.182 \\ 
              \textit{ethernet} & -9.83 & -213.09 & 6435.68 & 23.480 & 7.433 \\ 
              \textit{ecg} & -5.04 & -205.58 & 5873.96 & 22.760 & 5.765 \\ 
              \textit{linkruncca} & -41.40 & -146498.44 & 10380.14 & 28.330 & 12.290 \\ 
              \textit{xge\_mac} & -5.86 & -43.86 & 12747.57 & 7.556 & 14.090 \\ 
              \textit{fft256} & -17.14 & -550.53 & 26638.81 & 72.630 & 31.600 \\ \hline\hline
                Ratio Avg. & 100.00\% & 100.00\% & 100.00\% & 100.00\% & 100.00\% \\ \hline
            \end{tabular}
            \footnotesize \tnote{1} Dynamic Power. \tnote{2} Static Power.
        \end{threeparttable}
    }
\end{table}
\begin{table*}[tb!]
    \centering
    \footnotesize
    \caption{Five-run average PPA comparison with \textbf{WNS} as the primary objective and TNS/area/power as secondary metrics.}
    \label{tab:main_comparison}
    \renewcommand{\arraystretch}{1.1}
    \tabcolsep=3pt
    \resizebox{\linewidth}{!}
    { 
        \begin{threeparttable}
            \begin{tabular}{|c|ccc||ccc||ccc||ccc||ccc||}
              \hline
              \multirow{2}{*}{Benchmarks} &
              \multicolumn{3}{c||}{\textbf{WNS} (ps)} &
              \multicolumn{3}{c||}{TNS (ps)} &
              \multicolumn{3}{c||}{Area ($\mu$m$^2$)} &
              \multicolumn{3}{c||}{Dynamic Power (mW)} &
              \multicolumn{3}{c||}{Static Power ($\mu$W)}\\
              \cline{2-16}
             & ~\cite{zheng2025chatls}	& ~\cite{liu2024cbtune}	& SynAct	& ~\cite{zheng2025chatls}	& ~\cite{liu2024cbtune}	& SynAct	& ~\cite{zheng2025chatls}	& ~\cite{liu2024cbtune}	& SynAct & ~\cite{zheng2025chatls}	& ~\cite{liu2024cbtune}	& SynAct  & ~\cite{zheng2025chatls}	& ~\cite{liu2024cbtune}	& SynAct\\ \hline\hline
            \textit{uart16550} & -6.62 & -2.04 & \textbf{-1.18} & -6.69 & -2.04 & \textbf{-1.30} & 403.57 & \textbf{393.19} & 397.73 & 1.575 & \textbf{1.570} & 1.573 & 0.394 & \textbf{0.384} & 0.391 \\ 
            \textit{picorv32} & -11.92 & -2.94 & \textbf{0.06} & -329.10 & -268.32 & \textbf{0.00} & \textbf{1123.33} & 1146.73 & 1123.56 & \textbf{3.489} & 3.524 & 3.510 & \textbf{1.190} & 1.287 & 1.276 \\ 
            \textit{yacc} & -12.42 & -18.61 & \textbf{-4.38} & -595.46 & -996.20 & \textbf{-68.27} & \textbf{1155.38} & 1155.90 & 1159.37 & \textbf{4.115} & 4.162 & 4.150 & 1.306 & 1.333 & \textbf{1.304} \\ 
            \textit{aes\_core} & -9.83 & -3.68 & \textbf{-3.41} & -869.29 & \textbf{-108.98} & -118.37 & 1249.87 & 1210.61 & \textbf{1194.67} & 4.020 & 3.765 & \textbf{3.703} & 1.390 & 1.369 & \textbf{1.316} \\ 
            \textit{wb2axip} & -9.23 & -16.22 & \textbf{-6.78} & -4170.73 & -8771.95 & \textbf{-3488.09} & 1393.82 & 1523.99 & \textbf{1328.59} & 5.498 & 5.573 & \textbf{5.162} & 1.222 & 1.687 & \textbf{1.173} \\ 
            \textit{wb\_conmax} & -12.20 & -11.12 & \textbf{0.01} & -3562.51 & -4704.21 & \textbf{0.00} & 1700.23 & 1686.60 & \textbf{1684.22} & 4.141 & 4.392 & \textbf{4.040} & 1.368 & 1.469 & \textbf{1.314} \\ 
            \textit{arm9} & -8.25 & -8.44 & \textbf{-3.70} & -120.85 & -149.88 & \textbf{-37.42} & 2022.95 & \textbf{1971.41} & 2038.79 & 1.812 & \textbf{1.785} & 1.819 & 2.059 & \textbf{1.983} & 2.100 \\ 
            \textit{sha3} & -14.33 & -6.32 & \textbf{-3.82} & -8183.41 & -1849.45 & \textbf{-705.50} & 4079.15 & 2901.76 & \textbf{2837.75} & 30.540 & \textbf{15.380} & 17.210 & 3.979 & 3.261 & \textbf{2.761} \\ 
            \textit{spimaster} & -13.51 & -4.47 & \textbf{-3.04} & -142.14 & \textbf{-35.01} & -53.20 & 5448.90 & 5428.24 & \textbf{5415.65} & 11.280 & 11.270 & \textbf{11.240} & 6.165 & 6.166 & \textbf{6.128} \\ 
            \textit{ethernet} & -3.56 & -6.53 & \textbf{-2.85} & -304.41 & -67.81 & \textbf{-25.05} & 6438.15 & 6449.11 & \textbf{6406.14} & 23.620 & \textbf{23.280} & 23.320 & \textbf{7.080} & 7.465 & 7.378 \\ 
            \textit{ecg} & -3.35 & -4.56 & \textbf{-2.23} & \textbf{-22.11} & -453.69 & -33.09 & \textbf{5838.97} & 6304.30 & 5842.29 & \textbf{22.650} & 23.960 & 22.680 & \textbf{5.714} & 7.089 & \textbf{5.714} \\ 
            \textit{linkruncca} & -6.40 & \textbf{-3.14} & -6.19 & -12595.18 & \textbf{-801.79} & -5865.33 & \textbf{9832.23} & 9950.35 & 9876.56 & \textbf{27.390} & 27.770 & 27.660 & \textbf{11.520} & 11.870 & 11.730 \\ 
            \textit{xge\_mac} & -6.30 & -7.04 & \textbf{-4.36} & -34.54 & -35.47 & \textbf{-25.77} & \textbf{12613.77} & 12622.96 & 12668.15 & \textbf{7.478} & 7.540 & 7.541 & 13.970 & \textbf{13.960} & 14.010 \\ 
            \textit{fft256} & -12.68 & -34.27 & \textbf{-7.90} & -1805.46 & -1794.78 & \textbf{-137.37} & 26628.94 & \textbf{26475.57} & 26617.73 & \textbf{72.660} & 72.700 & \textbf{72.660} & 31.590 & \textbf{31.010} & 31.570 \\  \hline \hline
            Ratio Avg.\tnote{1} & 71.73\% & 66.67\% & \textbf{27.03\%} & 96.43\% & 77.14\% & \textbf{17.86\%} & 103.14\% & 101.02\% & \textbf{99.28\%} & 105.33\% & 99.79\% & \textbf{98.92\%} & 101.67\% & 105.51\% & \textbf{98.64\%} \\ \hline
            \end{tabular}
            \begin{tablenotes}
            \footnotesize
            \item[1] ``Ratio Avg.'' averages per-design metric ratios relative to bootstrap synthesis; lower values indicate greater improvement.
        \end{tablenotes}
        \end{threeparttable}
    }
\end{table*}
\begin{table}[tb!]
    \centering
    \footnotesize
    \caption{Runtime and token usage averaged over five runs.}
    \label{tab:runtime}
    {{{
    \renewcommand{\arraystretch}{1.1}
    \resizebox{0.97\linewidth}{!}
    { 
        \begin{threeparttable}
            \begin{tabular}{|c|cccc|cc|}
              \hline
              \multirow{2}{*}{Benchmark}  & \multicolumn{4}{c|}{Total execution time (s)} &  \multicolumn{2}{c|}{Token usage} \\ \cline{2-7}
              & Boot.\tnote{1} &~\cite{zheng2025chatls} &~\cite{liu2024cbtune} & SynAct  & Anal.\tnote{2} &  Opt.\tnote{3} \\ \hline \hline
        \textit{uart16550} & 19.02 & 187.93 & 428.59 & 425.20 & 43381 & 95823 \\ 
        \textit{picorv32} & 83.86 & 225.47 & 1617.28 & 783.49 & 43747 & 96444 \\ 
        \textit{yacc} & 101.57 & 225.55 & 2232.63 & 898.45 & 42262 & 96144 \\ 
        \textit{aes\_core} & 93.88 & 351.18 & 2478.49 & 1078.54 & 43955 & 95322 \\ 
        \textit{wb2axip} & 84.40 & 265.93 & 2323.97 & 822.28 & 45103 & 95840 \\ 
        \textit{wb\_conmax} & 121.26 & 258.71 & 3427.69 & 1383.27 & 44795 & 95416 \\ 
        \textit{arm9} & 617.47 & 980.17 & 10229.67 & 6941.71 & 45002 & 96493 \\ 
        \textit{sha3} & 248.03 & 515.75 & 5218.05 & 3242.30 & 43580 & 95474 \\ 
        \textit{spimaster} & 325.07 & 594.11 & 7005.90 & 2084.52 & 36585 & 95186 \\ 
        \textit{ethernet} & 327.66 & 848.57 & 6408.45 & 2706.38 & 51182 & 94415 \\ 
        \textit{ecg} & 253.50 & 567.90 & 5083.02 & 1850.21 & 37951 & 96071 \\ 
        \textit{linkruncca} & 1219.77 & 3379.13 & 24823.51 & 10079.57 & 42836 & 95464 \\ 
        \textit{xge\_mac} & 381.86 & 866.17 & 7669.93 & 1573.92 & 40991 & 96447 \\ 
        \textit{fft256} & 1551.27 & 2169.11 & 29751.55 & 10167.44 & 41060 & 96651 \\ \hline\hline
        Ratio Avg. & 100.00\% & 289.83\% & 2174.18\% & 988.62\% & - & - \\ \hline

            \end{tabular}
            \footnotesize \tnote{1} Bootstrap synthesis. \tnote{2} Analysis Agent. \tnote{3} Optimization Agent. 
        \end{threeparttable}
    }
    }}}
\end{table}

\minisection{Comparison setup and method reproduction}
\Cref{tab:original_statistics} reports the PPA metrics after the mandatory bootstrap synthesis, which configures the technology library and timing constraints, reads and elaborates each RTL design, and runs the default \texttt{optimize} command without additional recipe-level tuning. These metrics serve as the reference for subsequent comparisons.
The ``Ratio Avg.'' is the mean of the per-design ratios between each method and bootstrap synthesis. Each ratio is computed by dividing the method result by its bootstrap value. For WNS and TNS, the reported value is first converted to its violation magnitude, $\max(0,-\mathrm{slack})$, so timing closure contributes zero. A lower ``Ratio Avg.'' indicates greater improvement.

\Cref{tab:main_comparison} compares two baselines representing distinct paradigms of synthesis optimization.
We set SynAct's iteration limit to $n_{\mathrm{iter}}=5$. To control optimization depth, all methods commit five actions, each consisting of a command or command sequence. This equalizes the committed-action budget but not the oracle-query budget, because the methods require different numbers of candidate evaluations, as detailed below. We therefore report the end-to-end runtime in \Cref{tab:runtime} alongside the PPA results.

ChatLS~\cite{zheng2025chatls} is an LLM-based end-to-end script customization method that combines multimodal RAG and chain-of-thought reasoning. As its one-shot script cannot use feedback from intermediate synthesis results, it provides a natural counterpart to SynAct's iterative optimization.
Since its source code is unavailable, we reproduce ChatLS following the published methodology. Under its original seven-benchmark 45~nm FreePDK setup, our reproduction matches every reported result within 5\%. We then port it without further tuning to our 14-design setup using ASAP7 and \altisyn, retaining the same prompts, retrieval configuration, and five-step script. The resulting comparison represents our controlled reproduction rather than the official implementation.

CBTune~\cite{liu2024cbtune} is a LinUCB-based contextual bandit method originally developed for synthesis sequence generation in \tool{ABC}~\cite{brayton2010abc}. We reproduce it using its public implementation.\footnote{\url{https://github.com/sallyliu921/CB-EVO}} Unlike ChatLS and SynAct, LinUCB requires predefined, enumerable arms for reward estimation and action selection, so adapting CBTune to \altisyn requires bounding the tool's much larger command space. We manually construct the seven entries in \Cref{tab:cbtune_action_space} from documented \altisyn commands and the optimization categories used by CBTune on \tool{ABC}. They cover timing optimization, mapping, sizing, phase assignment, retiming, and resynthesis, and remain fixed across all 14 benchmarks without design-specific tuning. At each of five steps, CBTune evaluates the arms, updates LinUCB using short- and long-term rewards, and applies the selected arm.

\begin{table}[tb!]
    \centering
    \caption{Bounded action space for CBTune reproduction.}
    \label{tab:cbtune_action_space}
    \scriptsize
    \renewcommand{\arraystretch}{0.92}
    \begin{tabular}{@{}c@{\hspace{3pt}}>{\ttfamily\raggedright\arraybackslash}p{0.90\columnwidth}@{}}
        \toprule
        \normalfont ID & \normalfont Command \\
        \midrule
        1 & optimize -mode timing -map\_effort high \\
        2 & tdopt -flat -perturb -stop \{1\} -area 1.0; phase\_assign -mode area\_recovery \\
        3 & optimize -mode timing -map\_effort high -incremental \\
        4 & tdopt -flat -resyn2 -area 1.000000 \\
        5 & retime\_register -period 0 -effort medium; optimize -incremental \\
        6 & impl\_select -mode down -scope s+ -tns; resize -mode normal -cost c>s>a -flat -area 1.0 \\
        7 & pre\_td; phase\_assign -flat -mode timing -effort medium; tdopt -flat -stop \{10 1.0\} -area 1.000000 \\
        \bottomrule
    \end{tabular}
\end{table}

SynAct performs five sequential iterations without a fixed action list, enabling it to reason over a broader command space than CBTune. In each iteration, the Analysis Agent diagnoses the latest synthesis reports, and the Optimization Agent combines this diagnosis with GraphRAG-retrieved documentation and BO-based experience to generate ten candidates. After synthesis evaluation, SynAct commits the best safe command to optimize the circuit and feeds the updated reports into the next iteration. Both agents use DeepSeek V3.1.

\minisection{Optimization results with WNS as the primary objective}
As shown in \Cref{tab:main_comparison}, with \textbf{WNS} as the primary optimization objective, SynAct achieves the best overall timing improvement among all methods.
In terms of ``Ratio Avg.'', SynAct reduces the remaining WNS violation to 27.03\% of that from bootstrap synthesis, substantially outperforming ChatLS~\cite{zheng2025chatls} (71.73\%) and CBTune~\cite{liu2024cbtune} (66.67\%). It achieves nonnegative WNS on two designs, whereas neither baseline does.
Notably, for \textit{wb2axip}, SynAct achieves WNS of -6.78\,ps versus -9.23\,ps (ChatLS) and -16.22\,ps (CBTune); for \textit{arm9}, it reaches -3.70\,ps versus -8.25\,ps and -8.44\,ps, respectively. For \textit{picorv32} and \textit{wb\_conmax}, SynAct further approaches timing closure with WNS values of 0.06\,ps and 0.01\,ps.

For secondary metrics, TNS is assigned lower priority than WNS and tends to improve as WNS becomes tighter.
SynAct reduces TNS to 17.86\% of the bootstrap synthesis result, compared with 96.43\% for ChatLS and 77.14\% for CBTune.
In terms of area and power trade-offs, SynAct keeps metrics close to the bootstrap synthesis levels, with slight reductions in area and power (99.28\% area, 98.92\% dynamic power, 98.64\% static power).
These results show that SynAct achieves substantial improvements in both WNS and TNS while maintaining overall PPA balance and efficiency.

\minisection{Runtime and LLM API token cost}
\Cref{tab:runtime} summarizes the end-to-end execution time of all methods.
All methods begin with the same bootstrap synthesis stage, whose runtime is reported in the ``Boot.'' column and included in each method's total runtime. The percentages represent end-to-end runtime normalized to the bootstrap time.
ChatLS has the lowest normalized runtime, averaging 289.83\% of the bootstrap synthesis time, because it generates the script in a single pass. CBTune has the highest runtime at 2174.18\% due to its bandit-based exploration. SynAct strikes a favorable balance at 988.62\%, less than half of CBTune's runtime, while achieving the best WNS ratio of 27.03\%, compared with 71.73\% for ChatLS and 66.67\% for CBTune. The additional runtime relative to single-pass ChatLS comes from feedback-driven candidate evaluation, which enables iterative refinement.
\begin{figure}[tb!]
    \centering
    \includegraphics[width=1\linewidth]{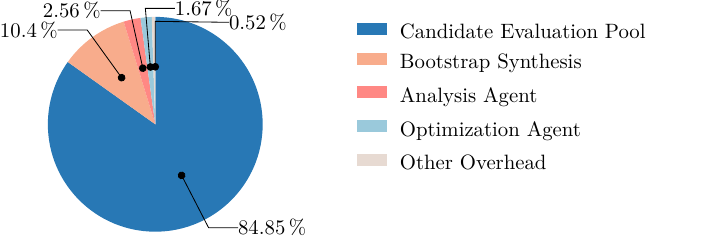}
    \caption{Runtime breakdown of SynAct.}
    \label{fig:runtime_pie}
\end{figure}

\Cref{fig:runtime_pie} further breaks down the internal runtime composition of SynAct.
Candidate evaluation dominates the runtime at 84.85\%. Each iteration evaluates ten LLM-generated candidates in parallel and commits the best safe action. This setting balances server capacity and robustness to LLM stochasticity: fewer candidates reduce synthesis cost, whereas more broaden exploration. Despite parallelism, repeated synthesis remains the primary bottleneck.
Bootstrap synthesis, the required initialization step, accounts for the second-largest share at 10.4\%. The Analysis and Optimization Agents account for 2.56\% and 1.67\%, respectively, while GrammarVAE encoding, decoding, and result processing account for the remaining 0.52\%.
Future work may consider early stopping upon WNS saturation or lightweight candidate filtering before execution to reduce unnecessary synthesis runs.

Token usage in \Cref{tab:runtime} is consistent across all 14 benchmarks. On average, SynAct consumes about 139k tokens per design, with 69\% from the Optimization Agent and 31\% from the Analysis Agent. The Optimization Agent uses more tokens because it processes more documents, even after RAG filtering. Total usage varies slightly from 132k to 146k, showing low cross-design variance and a stable ratio between the two agents.

\subsection{Optimization Trajectory Analysis}
\begin{figure}[tb!]
    \centering
    \includegraphics[width=1\linewidth]{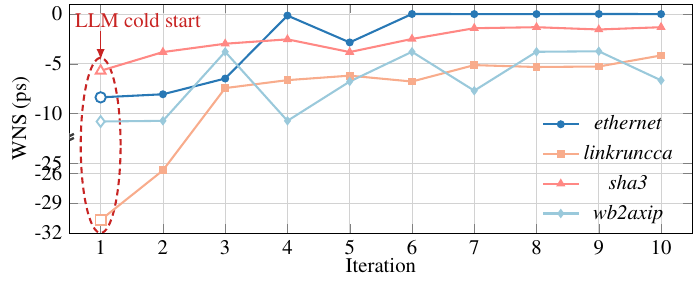}
    \caption{Optimization trajectory of SynAct.}
    \label{fig:case_study}
\end{figure}

Recall that SynAct incrementally optimizes a design over at most $n_{\mathrm{iter}}$ LLM-guided iterations, where $n_{\mathrm{iter}}$ is the tunable maximum number of iterations. We use $n_{\mathrm{iter}}=5$ in the main experiments in \Cref{tab:main_comparison} to balance optimization quality and runtime cost. \Cref{fig:case_study} evaluates the sensitivity to $n_{\mathrm{iter}}$ by plotting optimization trajectories from $n_{\mathrm{iter}}=1$ onward. For readability, we show four representative benchmarks: \textit{ethernet}, \textit{linkruncca}, \textit{sha3}, and \textit{wb2axip}. Their clear WNS improvements and well-separated curves provide a concise view of SynAct's overall optimization trend without the visual clutter of plotting all benchmarks.

The first iteration, labeled ``LLM cold start,'' refers to SynAct's initial attempt to generate optimization commands that rely almost solely on RAG-retrieved documents without historical experience, often leading to suboptimal results. As iterations proceed, SynAct refines its strategy using updated analyses and synthesis feedback, improving WNS over time. Most trajectories trend upward, while minor fluctuations in \textit{linkruncca} and \textit{sha3} reflect the exploratory nature of LLM-guided optimization. Although additional iterations can further improve WNS when runtime is less constrained, $n_{\mathrm{iter}}=5$ provides a practical balance between optimization quality and runtime cost, supporting our default setting. Overall, these results demonstrate the effectiveness of iterative adaptation driven by cumulative synthesis feedback.

\subsection{Generalizability Across LLMs}
To evaluate whether SynAct generalizes across different LLMs, we rerun the complete 14-design experiment using GPT-5.2 for both the Analysis and Optimization Agents under the same setup as \Cref{tab:main_comparison}. As reported in \Cref{tab:gpt_backend}, this configuration reduces the WNS and TNS Ratio Avg.\ to 20.37\% and 13.69\%, respectively, while keeping area and power within 1\% of the bootstrap results. The default DeepSeek V3.1 configuration already achieves a WNS Ratio Avg.\ of 27.03\% and substantially outperforms the baselines, confirming that SynAct's closed-loop diagnosis, retrieval, and experience refinement remain effective with different LLMs. GPT-5.2 further enhances report interpretation, diagnosis, and command generation within this pipeline, yielding additional timing gains. These results show that SynAct provides the adaptive optimization mechanism, while a stronger LLM can exploit it more effectively and raise the performance ceiling.

\begin{table}[tb!]
    \centering
    \footnotesize
    \caption{Five-run average PPA results of SynAct with GPT-5.2 under the same setup as \Cref{tab:main_comparison}.}
    \label{tab:gpt_backend}
    \renewcommand{\arraystretch}{1.1}
    \resizebox{\linewidth}{!}{
    \begin{threeparttable}
    \begin{tabular}{|c|ccccc|}
        \hline
        Benchmarks & \textbf{WNS}(ps) & TNS(ps) & Area ($\mu$m$^2$) & DP (mW)\tnote{1} & SP ($\mu$W)\tnote{2} \\ \hline\hline
        \textit{uart16550} & -1.43 & -1.53 & 401.39 & 1.584 & 0.395 \\
        \textit{picorv32} & 0.04 & 0.00 & 1131.68 & 3.542 & 1.303 \\
        \textit{yacc} & -3.28 & -59.29 & 1156.95 & 4.133 & 1.300 \\
        \textit{aes\_core} & -2.90 & -158.80 & 1207.75 & 3.747 & 1.345 \\
        \textit{wb2axip} & -6.46 & -3352.07 & 1329.51 & 5.163 & 1.175 \\
        \textit{wb\_conmax} & -0.59 & -9.60 & 1685.97 & 4.363 & 1.460 \\
        \textit{arm9} & -4.42 & -47.12 & 2034.10 & 1.817 & 2.077 \\
        \textit{sha3} & -2.54 & -246.83 & 2839.00 & 17.220 & 2.765 \\
        \textit{spimaster} & 0.02 & 0.00 & 5387.82 & 11.240 & 6.103 \\
        \textit{ethernet} & -2.94 & -17.15 & 6377.96 & 23.530 & 7.330 \\
        \textit{ecg} & 0.01 & 0.00 & 5846.38 & 22.690 & 5.722 \\
        \textit{linkruncca} & 0.00 & 0.00 & 9815.93 & 27.540 & 11.440 \\
        \textit{xge\_mac} & -4.36 & -25.63 & 12683.96 & 7.540 & 14.020 \\
        \textit{fft256} & -6.49 & -151.60 & 26612.28 & 72.650 & 31.560 \\
        \hline\hline
        Ratio Avg. & 20.37\% & 13.69\% & 99.36\% & 99.65\% & 99.41\% \\
        \hline
    \end{tabular}
    \footnotesize \tnote{1} Dynamic Power. \tnote{2} Static Power.
    \end{threeparttable}}
\end{table}

\subsection{Validation of Latent-Space BO Guidance}
\label{sec:latent_diagnostics}

To validate BO-guided experience refinement, we test two properties required by its design. First, if GrammarVAE provides a useful local search prior, commands that are close in latent space should produce more similar synthesis rewards than distant commands. Using $r(C)$ from \Cref{eq:reward}, we measure this similarity by $\Delta r(C_i,C_j)=|r(C_i)-r(C_j)|$, where a smaller value indicates more consistent optimization outcomes. As shown in \Cref{tab:latent_locality}, near command pairs have a mean $\Delta r$ of 0.188 versus 0.238 for far pairs, a 21.0\% reduction. Second, the BO guidance should remain useful after the LLM converts a decoded seed into tool-valid candidates. BO-seeded candidates achieve a mean reward of 0.297 versus 0.289 for RAG-grounded candidates and obtain the highest mean reward in 65.3\% of benchmark iterations. These results support the two design properties: locally consistent latent neighborhoods and BO guidance that remains effective after tool-aware LLM refinement.

\begin{table}[tb!]
    \centering
    \footnotesize
    \caption{Diagnostics of GrammarVAE latent locality and BO-seed utility over 14 benchmarks.}
    \label{tab:latent_locality}
    \renewcommand{\arraystretch}{1.1}
    \resizebox{0.92\linewidth}{!}{
    \begin{tabular}{|l|l|c|}
        \hline
        Category & Diagnostic & Result \\
        \hline\hline
        \multirow{2}{*}{Latent locality}
        & Mean $\Delta r$, near pairs & 0.188 \\
        & Mean $\Delta r$, far pairs & 0.238 \\
        \hline
        \multirow{4}{*}{BO seed utility}
        & BO-seeded mean reward & 0.297 \\
        & RAG-grounded mean reward & 0.289 \\
        & Iterations where BO has highest mean & 65.3\% \\
        & Iterations where best is BO-seeded & 47.2\% \\
        \hline
    \end{tabular}}
\end{table}

\begin{figure*}[tb!]
    \centering
    \includegraphics[width=\linewidth]{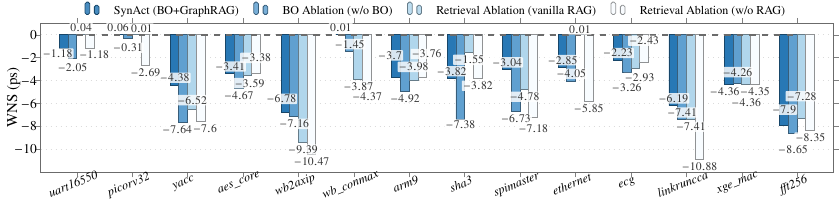}
    \caption{Ablation of SynAct components on WNS. The full configuration (BO + GraphRAG) achieves the best timing improvement.}
    \label{fig:graphrag_wns_delta}
\end{figure*}

\subsection{Ablation Study on BO and Retrieval Modules}
\label{sec:bo_retrieval_ablation}
\Cref{fig:graphrag_wns_delta} presents the ablation results using the full SynAct configuration (``BO+GraphRAG'') as the reference.
In this configuration, candidate generation combines report diagnosis, documentation retrieved by GraphRAG, and GrammarVAE seeds guided by BO.
All variants generate ten candidates per iteration, evaluate them through synthesis, and select the best safe command, thereby isolating the effects of BO and retrieval.
The ``w/o BO'' variant removes BO guidance and the reuse of historical commands, while ``vanilla RAG'' uses flat retrieval and ``w/o RAG'' provides the full documentation directly.

Without BO, timing performance clearly degrades, with the WNS Ratio Avg.\ rising from 27.0\% to 37.5\%, indicating less effective WNS improvement.
The full configuration outperforms this variant on 13 of 14 benchmarks, with WNS improving from -7.16\,ps to -6.78\,ps on \textit{wb2axip} and from -8.65\,ps to -7.90\,ps on \textit{fft256}.
These results show that BO-guided experience reuse steers exploration of the command space toward better optimization outcomes.

The ``vanilla RAG'' variant raises the WNS Ratio Avg.\ from 27.0\% to 28.6\%, while the ``w/o RAG'' variant further increases it to 38.3\%.
WNS degradations exceeding 2\,ps on \textit{yacc} and \textit{linkruncca} suggest that GraphRAG's structured representation captures richer contextual dependencies than flat retrieval or direct use of the full documentation.
Overall, SynAct achieves the most effective timing improvement among all variants, with BO supporting effective exploration and GraphRAG providing structured contextual knowledge for optimization.

%% file: doc/conclu.tex
\section{Conclusion}
\label{sec:conclusion}
In summary, SynAct is an adaptive closed-loop LLM framework for iterative PPA optimization in commercial synthesis, addressing the limitations of fixed-action search and one-shot script generation.
By combining synthesis feedback, prior experience, and retrieved tool knowledge, SynAct enables state-aware decisions with explicit rationales while improving timing and maintaining balanced area and power.
Although evaluated only on \altisyn and open-source designs, SynAct is designed for portability to other logic synthesis tools with scripting interfaces, command documentation, and structured PPA and timing reports by replacing its tool-specific adapters and knowledge base. Future work will validate cross-tool portability and industrial-scale performance and reduce candidate-evaluation overhead.

%% file: doc/acknowledge.tex
\section*{Acknowledgment}
The authors thank ZeniSyn Design Systems for providing access to \altisyn{} and related technical support. Regarding the use of generative AI, DeepSeek V3.1 and GPT-5.2 were used in SynAct to analyze synthesis reports and generate candidate commands, while OpenAI Codex was used solely for language editing and grammar checking. All technical content, ideas, and results are original work by the authors.

%% file: doc/bio.tex
\begin{IEEEbiography}
    [{\includegraphics[width=1in]{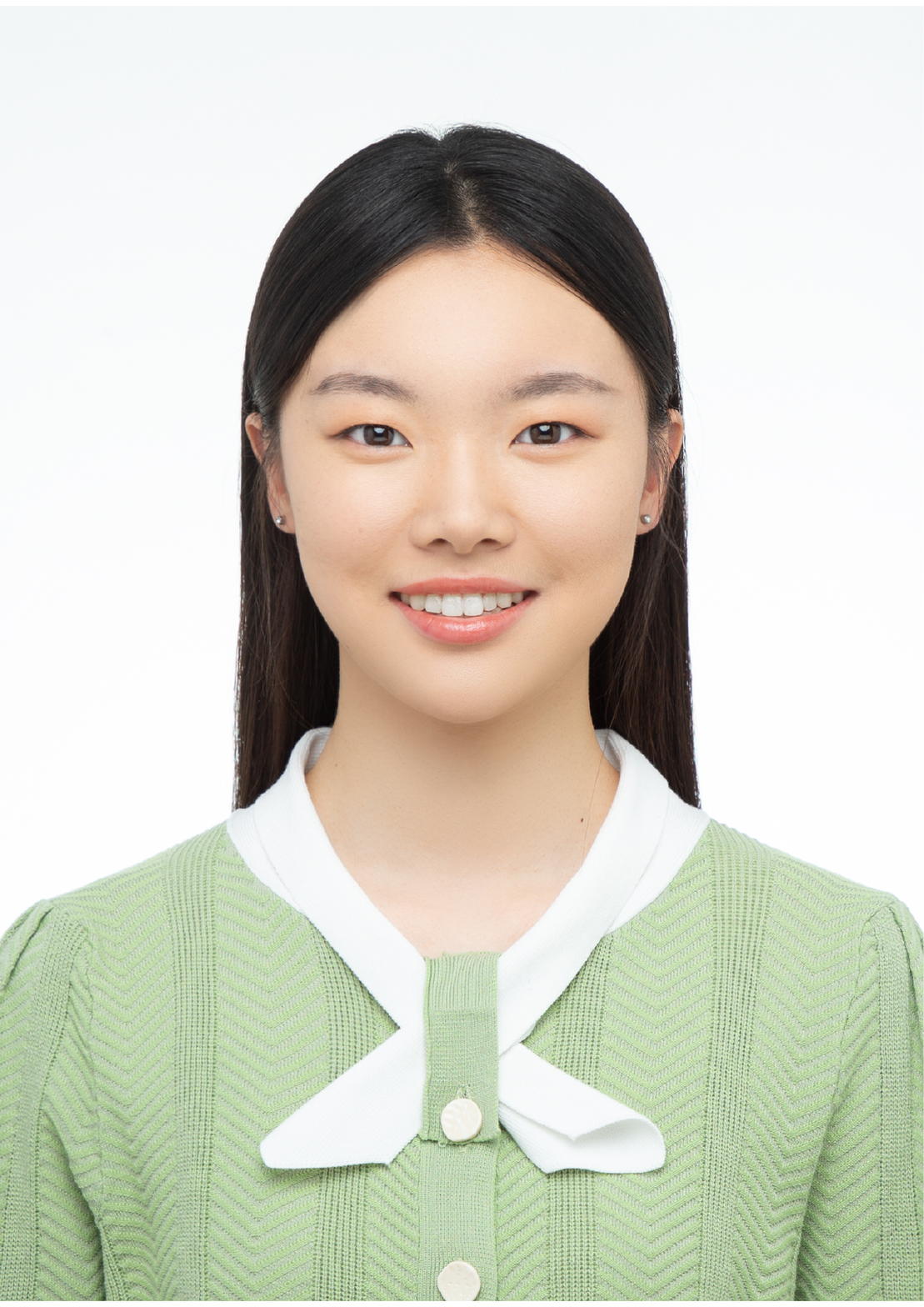}}]
    {Fangzhou Liu}
    received her B.Eng.~degree from the School of Electronic Science and Engineering, Nanjing University, in 2023. She is currently pursuing the Ph.D. degree at the Chinese University of Hong Kong under the supervision of Prof. Bei Yu since Fall 2023. Her research focuses on applying machine learning techniques to EDA and logic synthesis optimization.
\end{IEEEbiography}

\vspace{-.41in}
\begin{IEEEbiography}
    [{\includegraphics[width=1in]{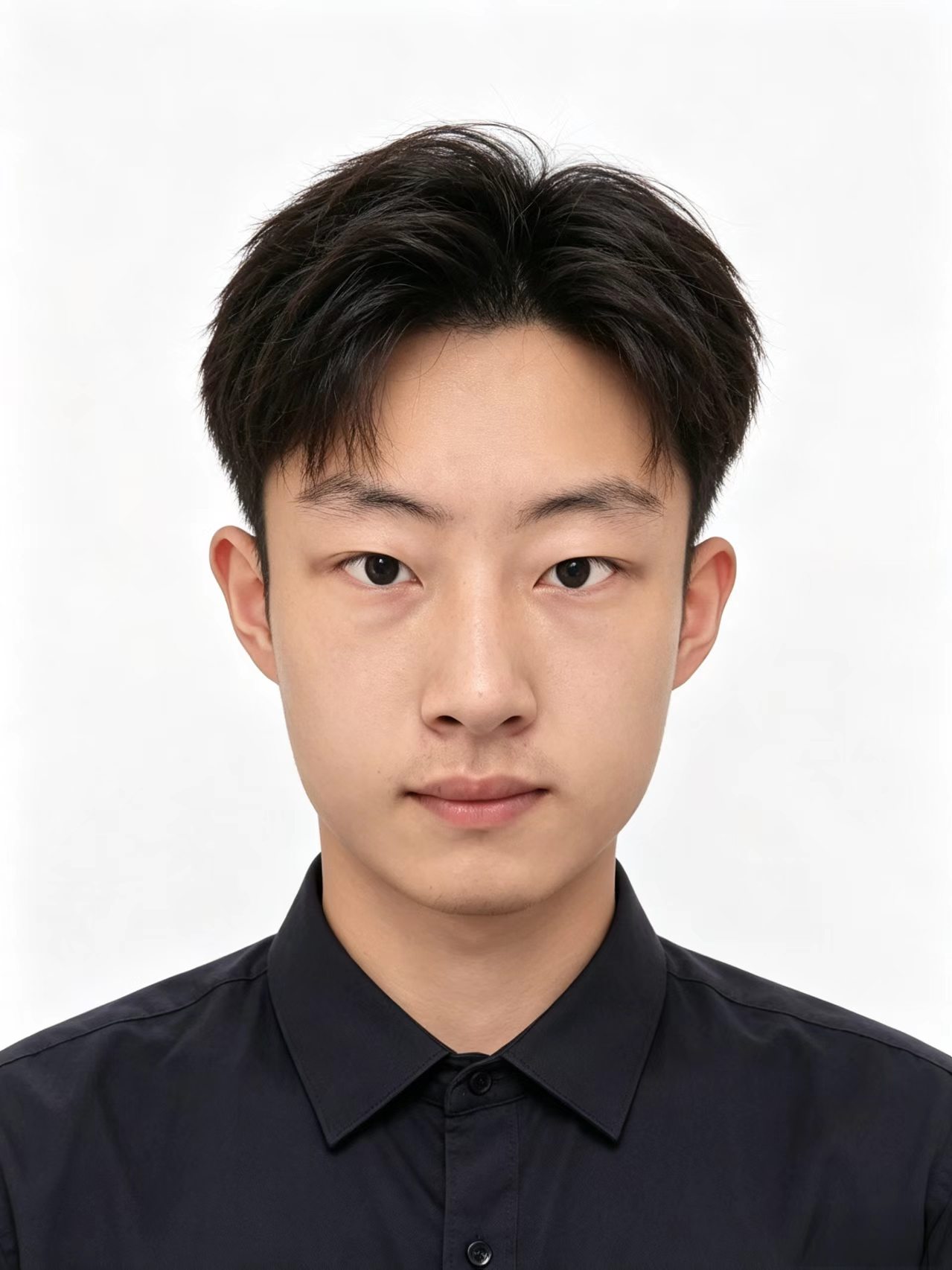}}]
    {Peiyi Han}
    received the B.E. degree in automation from Tsinghua University in 2022 and the M.S. degree from the Institute of Computing Technology, University of Chinese Academy of Sciences, in 2025, under the supervision of Prof. Yunji Chen. He is currently pursuing the Ph.D. degree with the Department of Computer Science and Engineering, The Chinese University of Hong Kong, under the supervision of Prof. Bei Yu. His research interests include large language models and machine learning, with applications in electronic design automation (EDA).
\end{IEEEbiography}

\vspace{-.41in}
\begin{IEEEbiography}
    [{\includegraphics[width=1in]{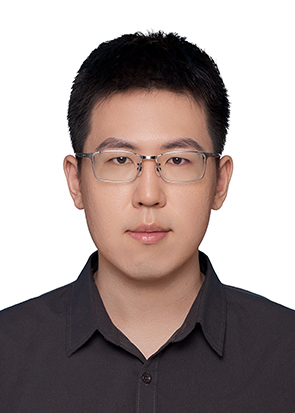}}]
    {Jiawei Liu}
    received his B.Eng. and Ph.D. degrees from Beijing University of Posts and Telecommunications in 2020 and 2025, respectively. He is currently a Postdoctoral Fellow with the Department of Computer Science and Engineering, The Chinese University of Hong Kong, under the supervision of Prof. Bei Yu. His research focuses on electronic design automation and artificial intelligence.
\end{IEEEbiography}

\vspace{-.41in}
\begin{IEEEbiography}
    [{\includegraphics[width=1in]{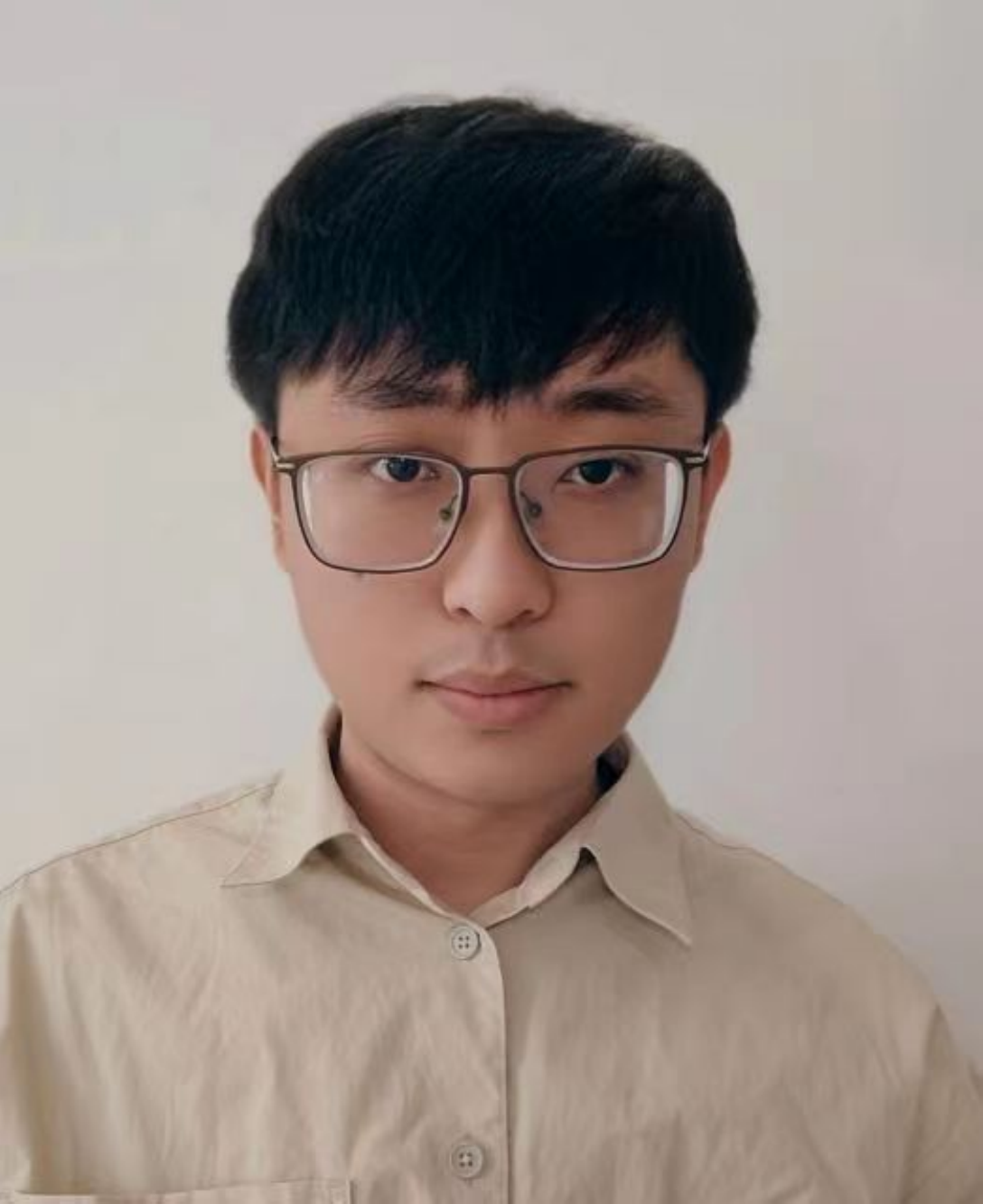}}]
    {Yuan Pu}
	received the B.S. degree in computer science from The Chinese University of Hong Kong, Hong Kong, in 2022.
He is now a third-year Ph.D. Student at the Department of Computer Science and Engineering, the Chinese University of Hong Kong (CUHK), supervised by Prof. Bei YU since 2023 Fall. His research interest includes Combinatorial Algorithms/AI in EDA.
\end{IEEEbiography}

\vspace{-.41in}
\begin{IEEEbiography}
    [{\includegraphics[width=1in,trim=9 0 9 0,clip]{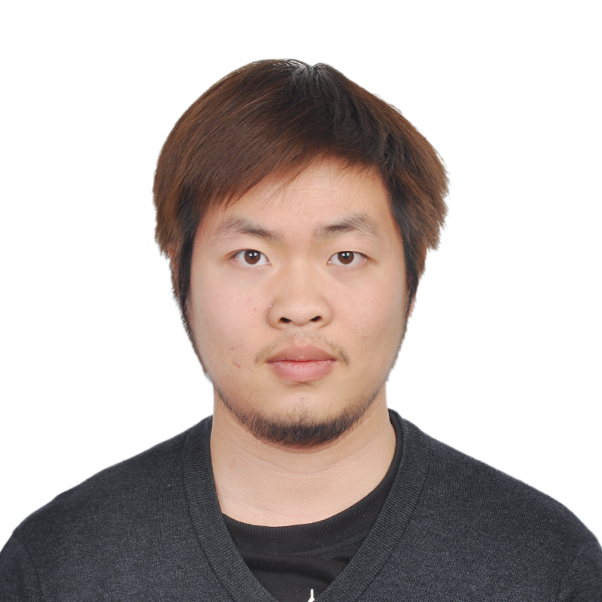}}]
    {Zhuolun He}
	 received the B.S. degree in computer science and engineering from Peking University, Beijing, China, in 2017, and the Ph.D. degree from the Chinese University of Hong Kong, Hong Kong, in 2023. He is an incoming assistant professor with Tsinghua University, Beijing, China.

\end{IEEEbiography}

\vspace{-.41in}
\begin{IEEEbiography}
    [{\includegraphics[width=1in]{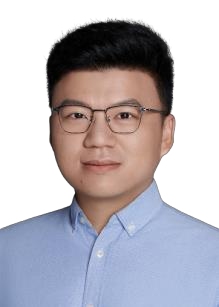}}]
    {Rongliang~Fu}
    received his Ph.D. in Computer Science and Engineering from The Chinese University of Hong Kong in January 2026, following an M.S. from the University of Chinese Academy of Sciences in June 2021 and a B.S. in Software Engineering from Northwestern Polytechnical University in June 2018. He has authored over 30 papers across major journals (IEEE TC and IEEE TCAD) and conferences(DAC, DATE, ICCAD, etc.). His research spans electronic design automation and EDA for superconducting electronics.
\end{IEEEbiography}

\vspace{-.41in}
\begin{IEEEbiography}
    [{\includegraphics[width=1in]{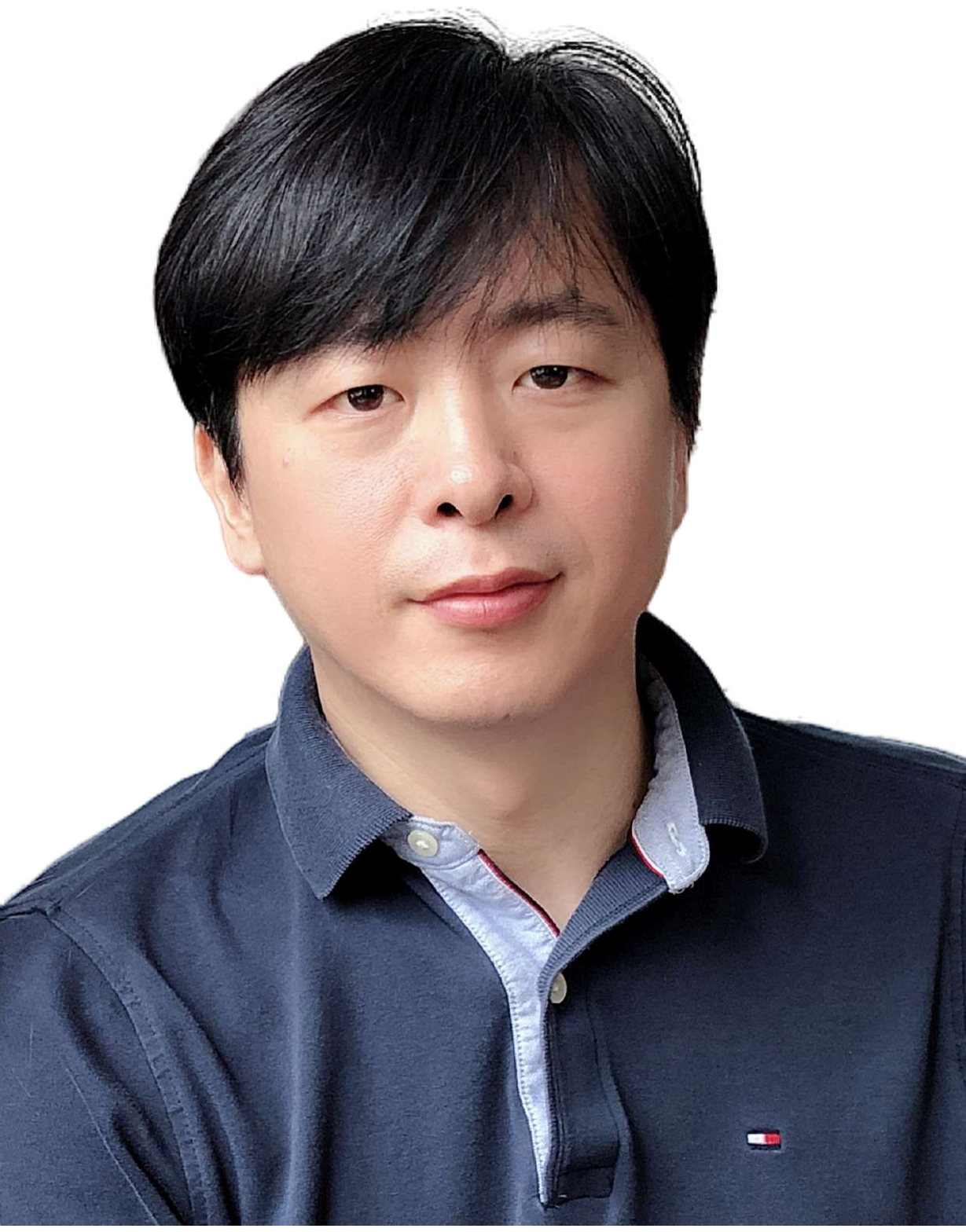}}]
    {Tsung-Yi Ho}
    (F'24)
    is a Professor in the Department of Computer Science and Engineering, The Chinese University of Hong Kong (CUHK). He received his Ph.D. in Electrical Engineering from National Taiwan University in 2005. His research interests include several areas of computing and emerging technologies, especially in the design automation of microfluidic biochips. He was a recipient of the Best Paper Award at the IEEE Transactions on Computer-Aided Design of Integrated Circuits and Systems in 2015. Currently, he serves as the VP Conferences of IEEE CEDA, and the Executive Committee of ASP-DAC and ICCAD. He is a Distinguished Member of ACM and a Fellow of IEEE.
\end{IEEEbiography}

\vspace{-.41in}
\begin{IEEEbiography}[{\includegraphics[width=1in]{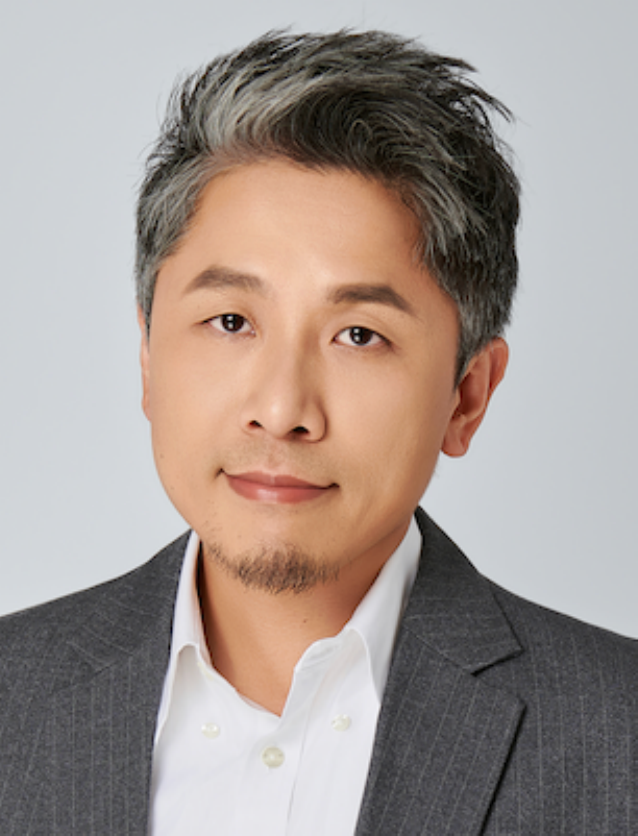}}]
    {Bei Yu}
    (M'15-SM'22)
    received the Ph.D.~degree from The University of Texas at Austin in 2014.
    He is currently a Professor in the Department of Computer Science and Engineering, The Chinese University of Hong Kong.
    He has served as TPC Chair of ACM/IEEE Workshop on Machine Learning for CAD, and in many journal editorial boards and conference committees.
    He received eleven Best Paper Awards from ICCAD 2024 \& 2021 \& 2013,
    IEEE TSM 2022, DATE 2022, ASPDAC 2021 \& 2012, ICTAI 2019, Integration, the VLSI Journal in 2018,
    ISPD 2017, SPIE Advanced Lithography Conference 2016, six ICCAD/ISPD contest awards,
    IEEE CEDA Ernest S. Kuh Early Career Award in 2021,
    DAC Under-40 Innovator Award in 2024,
    and Hong Kong RGC Research Fellowship Scheme (RFS) Award in 2024.
\end{IEEEbiography}